\documentclass[journal,twoside,web]{utils/ieeecolor}

\def\logoname{utils/LOGO-tuffc-web}
\definecolor{subsectioncolor}{rgb}{0.067,0.627,0.859}
\def\journalname{IEEE Transactions on Ultrasonics, Ferroelectrics, and Frequency Control}

\def\fulltitle{Hadamard-Based Recursive Aperture Decoded Ultrasound Imaging (READI) With Estimated Motion-Compensated Compounding ($\text{EMC}^2$) Using Top-Orthogonal to Bottom Electrode (TOBE) Arrays}

\usepackage{iftex}
\ifPDFTeX
  \usepackage[T1]{fontenc}
\fi
\usepackage{newtxtext,newtxmath}

\usepackage{cite}
\usepackage{graphicx}
\graphicspath{{figures/}}
\DeclareGraphicsExtensions{.pdf,.jpeg,.png}
\usepackage{amsmath}
\usepackage{amsfonts}
\usepackage{algorithmic}
\usepackage{array}
\usepackage[caption=false,font=footnotesize]{subfig}
\usepackage{url}
\usepackage{bm}
\usepackage{multirow}
\usepackage{dblfloatfix}
\usepackage{annotate-equations}
\usepackage{makecell}

\usepackage{acro}

\newcommand{\numberthis}{\addtocounter{equation}{1}\tag{\theequation}}

\makeatletter
\pretocmd{\@afterheading}{\color{black}\normalcolor}{}{}
\makeatother

\usepackage{textcomp}
\usepackage{wrapfig,colortbl}
\definecolor{abstractbg}{rgb}{1,0.969,0.914}
\def\BibTeX{{\rm B\kern-.05em{\sc i\kern-.025em b}\kern-.08em
    T\kern-.1667em\lower.7ex\hbox{E}\kern-.125emX}}

\newcommand{\sigmatt}[1]{\textbf{\textrm{#1}}}
\newcommand{\sigmatm}[1]{\bm{\mathrm{#1}}}

\newcommand{\emcTwo}{EMC\textsuperscript{2}}

\newcommand{\txgroup}{$\sigmatm{g'}_{s}(t)$}
\newcommand{\partDec}{$\sigmatm{D}(t)$}
\newcommand{\partDecGroup}{$\sigmatm{d}_{s}(t)$}

\DeclareAcronym{tobe}{
  short = TOBE,
  long  = Top-Orthogonal to Bottom Electrode
}

\DeclareAcronym{forces}{
  short = FORCES,
  long  = Fast Orthogonal Row-Column Electronic Scanning
}

\DeclareAcronym{readi}{
  short = READI,
  long  = Recursive Aperture Decoded Imaging
}

\DeclareAcronym{cmut}{
  short             = CMUT,
  short-plural-form = CMUTs,
  long              = Capacitive Micromachined Ultrasonic Transducer,
  long-plural-form  = Capacitive Micromachined Ultrasonic Transducers
}

\DeclareAcronym{rca}{
  short             = RCA,
  short-plural-form = RCAs,
  long              = Row-Column Array,
  long-plural-form  = Row-Column Arrays
}

\DeclareAcronym{sta}{
  short = STA,
  long  = Synthetic Transmit Aperture
}

\DeclareAcronym{emc2}{
  short = EMC\textsuperscript{2},
  long  = Estimated Motion-Compensated Compounding
}

\DeclareAcronym{vls}{
  short = VLS,
  long  = Virtual Line Source
}

\DeclareAcronym{tpw}{
  short = TPW,
  long  = Tilted Plane Wave
}

\DeclareAcronym{snr}{
  short = SNR,
  long  = Signal-to-Noise Ratio
}

\DeclareAcronym{cnr}{
  short = CNR,
  long  = Contrast-to-Noise Ratio
}

\DeclareAcronym{gcnr}{
  short = gCNR,
  long  = Generalized Contrast-to-Noise Ratio
}

\DeclareAcronym{das}{
  short = DAS,
  long  = Delay-and-Sum
}

\DeclareAcronym{ncc}{
  short = NCC,
  long  = Normalized Cross-Correlation
}

\DeclareAcronym{uforces}{
  short = uFORCES,
  long  = Ultrafast FORCES
}

\DeclareAcronym{cuda}{
  short = CUDA,
  long  = Compute Unified Device Architecture
}

\DeclareAcronym{npp}{
  short = NPP,
  long  = NVIDIA Performance Primitives
}

\DeclareAcronym{cufft}{
  short = cuFFT,
  long  = CUDA Fast Fourier Transform
}

\DeclareAcronym{cublas}{
  short = cuBLAS,
  long  = CUDA Basic Linear Algebra Subroutine
}

\DeclareAcronym{fft}{
  short             = FFT,
  short-plural-form = FFTs,
  long              = Fast Fourier Transform,
  long-plural-form  = Fast Fourier Transforms
}

\DeclareAcronym{prf}{
  short = PRF,
  long  = Pulse Repetition Frequency
}

\DeclareAcronym{pdf}{
  short             = PDF,
  short-plural-form = PDFs,
  long              = Probability Density Function,
  long-plural-form  = Probability Density Functions
}

\DeclareAcronym{fov}{
  short             = FOV,
  short-plural-form = FOVs,
  long              = Field of View,
  long-plural-form  = Fields of View
}

\DeclareAcronym{svd}{
  short = SVD,
  long  = Singular Value Decomposition
}

\DeclareAcronym{hercules}{
  short = HERCULES,
  long  = Hadamard Encoded Row Column Ultrasonic Expansive Scanning
}

\DeclareAcronym{bpm}{
  short = BPM,
  long  = beats per minute
}

\DeclareAcronym{cfw}{
  short = CFW,
  long  = Coherence Factor Weighting
}

\DeclareAcronym{psf}{
  short = PSF,
  short-plural-form = PSFs,
  long  = Point-Spread Function,
  long-plural-form = Point-Spread Functions
}

\begin{document}

\title{\fulltitle}

\author{Tyler~Keith~Henry,~\IEEEmembership{Graduate~Student~Member,~IEEE,}
        Darren~Dahunsi,~\IEEEmembership{Graduate~Student~Member,~IEEE,}
        Randy~Palamar,~\IEEEmembership{Graduate~Student~Member,~IEEE,}
        Negar~Majidi,~\IEEEmembership{Graduate~Student~Member,~IEEE,}
        Ying~Wan,~\IEEEmembership{Graduate~Student~Member,~IEEE,}
        Mohammad~Rahim~Sobhani,~\IEEEmembership{Member,~IEEE,}
        Afshin~Kashani~Ilkhechi,~\IEEEmembership{Member,~IEEE,}
        and~Roger~Zemp,~\IEEEmembership{Member,~IEEE}
\thanks{Funding for this work was provided by the National Institutes of Health (1R21HL161626-01 and EITTSCA R21EYO33078), Alberta Innovates (AICE 202102269, CASBE 212200391 and LevMax 232403439), MITACS (IT46044 and IT41795), CliniSonix Inc., NSERC (2025-05274), the Alberta Cancer Foundation and the Mary Johnston Family Melanoma Grant (ACF JFMRP 27587), the Government of Alberta Cancer Research for Screening and Prevention Fund (CRSPPF 017061), an Innovation Catalyst Grant to MRS, and INOVAIT (2023-6359). We are grateful to the nanoFAB staff at the University of Alberta for facilitating array fabrication. \textit{Corresponding authors: Tyler Keith Henry (email: tkhenry@ualberta.ca), Roger Zemp (email: rzemp@ualberta.ca).}}%
\thanks{The authors are with the Department of Electrical and Computer Engineering, University of Alberta, Edmonton, AB, T6G 1H9, Canada}
}

\IEEEtitleabstractindextext{%
\fcolorbox{abstractbg}{abstractbg}{%
\begin{minipage}{\textwidth}\rightskip2em\leftskip\rightskip\bigskip
\begin{wrapfigure}[11]{r}{3in}%
\vspace{-1pc}\hspace{-3pc}\includegraphics[width=3in]{figures/drawings/graphical_abstract.pdf}
\end{wrapfigure}%
\begin{abstract}
Hadamard matrix-based aperture encoding is a method for producing synthetic aperture datasets with high \acl{snr}s. Recently, the pulse inversion capabilities of bias-sensitive \ac{tobe} arrays have driven the development of multiple Hadamard-based sequences. These sequences produce high-quality static images but are sensitive to motion. This work introduces \ac{readi} and \ac{emc2}, which look to reduce this sensitivity. \acs{readi} is a novel decoding and beamforming technique for Hadamard aperture-encoded sequences that produces multiple low-resolution images from subsets of the full sequence. These \acs{readi} images are less affected by motion and sum to form the complete high-resolution image. \acs{emc2} describes the process of comparing these low-resolution images to estimate the underlying motion, then warping them to align before compounding. This produces a high-resolution image that is resiliant to motion. \acs{readi} with \acs{emc2} applied to the TOBE-based \ac{forces} sequence. It is shown to fully restore images corrupted by probe motion and to recover tissue speckle and boundaries in images of a beating heart phantom. \acs{readi} low-resolution images by themselves are demonstrated to be a marked improvement over a sparse Hadamard scheme with the same transmit count, and are able to recover blood speckle at a flow rate of 42 cm/s. 
\end{abstract}

\bigskip
\end{minipage}}}

\maketitle
\begin{table*}[!t]
\arrayrulecolor{subsectioncolor}
\setlength{\arrayrulewidth}{1pt}
{\sffamily\bfseries\begin{tabular}{lp{6.75in}}\hline
\rowcolor{abstractbg}\multicolumn{2}{l}{\color{subsectioncolor}{\itshape
Highlights}{\Huge\strut}}\\
\rowcolor{abstractbg}$\bullet$ & Hadamard-encoded sequences can be recursively decoded into subgroups with fewer motion artifacts. These subgroups form low-resolution images that can be coherently compounded to form a high-resolution image.\\
\rowcolor{abstractbg}$\bullet${\large\strut} & Applying motion compensation before compounding yields a motion-robust high-resolution image.\\
\rowcolor{abstractbg}$\bullet${\large\strut} & This allows for ultrafast imaging with reduced motion artifacts and improved flow detection using only B-mode images.\\[2em]\hline
\end{tabular}}
\setlength{\arrayrulewidth}{0.4pt}
\arrayrulecolor{black}
\end{table*}

\section{Introduction} \label{sec:Introduction}

\acreset{forces}
\acreset{tobe}
\acreset{snr}
\acreset{readi}
\acreset{emc2}

\IEEEPARstart{S}{ynthetic} Transmit Aperture (STA) ultrasound imaging is a well-established alternative to traditional scan-line imaging that offers global transmit and receive focusing \cite{sa_review_jensen}. \acs{sta} combines a series of unfocused pulses from small sub-apertures into a large "synthetic" aperture spanning the entire array. The diverging transmissions maximize image width, while frame rate can be increased using sparse transmit schemes \cite{highspeed_div_sa} \cite{sparse_3d_sa}. 

\ac{forces} \cite{forces_ceroici} \cite{forces_comp_preprint} is a recently developed 2D \ac{sta} imaging technique that utilizes Hadamard aperture encoding with bias-sensitive \ac{tobe} arrays \cite{tobe_cmut} to achieve the same synthetic aperture as \ac{sta} while transmitting with the complete array on each event. This maximizes the total energy transmitted into the volume, boosting \ac{snr} and penetration of all signals.

\begin{figure*}
    \centering
    \includegraphics[width=\linewidth]{figures/drawings/tobe_hardware_combined.pdf}
    \caption{Operation of a relaxor-based TOBE array. A) Physical layout of elements and channels. B) Row-Column array functionality achieved by biasing all elements uniformly. C) Unique TOBE array functionality enabled by bias control. D) Hardware block diagram for a TOBE array and ultrasound system.}
    \label{fig:tobe_array}
\end{figure*}

Motion in the imaged volume during an \ac{sta} sequence can prevent the transmits from coherently compounding and lead to significant image blur \cite{sa_motion} \cite{sa_review_jensen}. This effect is exacerbated by \ac{forces}, as motion also prevents proper decoding of the aperture. Sparse \ac{sta} transmit schemes can increase the frame rate and reduce motion artifacts; this has been applied to \ac{forces} \cite{uforces} but comes at the cost of \ac{snr}, the value \ac{forces} is designed to maximize. 

This work introduces \ac{readi}, a new decoding and beamforming technique for Hadamard aperture-encoded sequences such as \ac{forces}. \ac{readi} separates the encoded dataset into sequential groups of transmit events, partially decodes them using a smaller Hadamard matrix, and then beamforms each group into a low-resolution image. These READI images are individually less corrupted by motion and, when combined, form the complete image at its full resolution and \ac{snr}. The underlying motion can be determined by comparing these \ac{readi} images and reversed by warping them. They can then form the complete image with improved motion resilience. We refer to this new process as \ac{emc2}. These methods were developed for and tested on the \ac{forces} sequence, but in principle can be applied to any Hadamard aperture-encoded sequence.

The paper includes cross-plane/3D renderings and videos of a human-sized beating-heart phantom with realistic motion and heart rates. The results indicate the potential for cross-plane imaging of the entire human heart using bias-switchable \ac{tobe} arrays, where most previous \acp{rca} have been limited to 3D imaging below the shadow of the aperture \cite{forces_comp_preprint}. It demonstrates robustness to motions encountered in future clinical scenarios and offers a combination of image quality, frame rate, and field of view not previously achieved with \ac{rca} methods.

Five experiments were conducted to evaluate the proposed methods. First, a \ac{forces} acquisition was simulated using Field II \cite{field_ii_1} \cite{field_ii_2} to validate that compounded \ac{readi} images perfectly reproduce the \ac{forces} image. The resolution of \ac{readi} images of various group sizes is also compared, along with the relative effects of motion. Second, a static anechoic cyst phantom was imaged with \ac{readi} and compared with \ac{uforces} \cite{uforces}, the aforementioned sparse version of \ac{forces}. Then, the same phantom was imaged while the probe was in motion to determine if \ac{emc2} can recover the static image. Next, a beating heart phantom was imaged in motion and compensated with \ac{emc2} to evaluate performance on non-uniform tissue motion. Finally, a flow phantom was imaged to determine if \ac{readi} images can resolve blood flow and speckle.

\section{Background} \label{sec:Background}
This section will provide a high-level overview of \ac{tobe} arrays and their unique capabilities, followed by a detailed review of \ac{forces}. It will cover how \ac{forces} improves upon \ac{sta} with Hadamard encoding, how the dataset is beamformed, and the unique benefits of \ac{forces} imaging with \ac{tobe} arrays, including multi-plane imaging. The equations introduced in section \ref{sec:forces_background} are necessary for the derivation of \ac{readi} in section \ref{sec:methods}.

\subsection{\Ac{tobe} Arrays}
\acf{tobe} arrays \cite{tobe_cmut} are a refinement of \aclp{rca} \cite{rca_sa} \cite{rca_composite} \cite{rca_functional} that allow for individual element addressing while retaining the same channel count and geometry. In a standard \ac{rca}, elements are organized into an $N \times M$ grid, where each element is connected to one row and one column contact. When transmitting or receiving along a contact, all connected elements transmit simultaneously and contribute equally to the received signal. \ac{tobe} arrays eliminate this coupling by using electrostrictive relaxor materials, which exhibit a piezoelectric response that is linearly proportional to an external DC bias \cite{relaxors} \cite{forces_ceroici_3d_comparison}. In the absence of this bias, the elements will not transmit or receive signals, while two equal and oppositely biased elements will have inverted responses. Figure \ref{fig:tobe_array} highlights the construction and biasing of a \ac{tobe} array, along with a complete block diagram of the imaging system with custom DC bias electronics. This work uses relaxor-based TOBE arrays; however, they can also be constructed from Capacitive Micromachined Ultrasonic Transducers (CMUTs) \cite{tobe_cmut}.

If all elements are biased with a fixed voltage, then \ac{tobe} arrays act identically to traditional \acp{rca} (Figure \ref{fig:tobe_array}B) and are capable of performing the same imaging sequences \cite{forces_comp_preprint}. Additionally, their bias-modulation capabilities (Figure \ref{fig:tobe_array}C) make them well-suited to new methods such as Hadamard aperture encoding. This spatial encoding requires some elements to transmit inverted pulses, which are difficult to implement on many clinical systems without modification \cite{delay_hadamard_encoding} \cite{s_matrix_encoding}. \ac{tobe} arrays can instead linearly invert the pulse by applying a negative bias to the transmitting element(s) as demonstrated in Figure \ref{fig:tobe_array}C. Multiple \ac{tobe} imaging sequences have been implemented with this encoding, such as \ac{forces} \cite{forces_ceroici} and \ac{hercules} \cite{hercules_preprint}. 

\ac{tobe} arrays enable individual-element addressing with a significantly reduced channel count compared to fully wired matrix arrays. Current commercially available matrix arrays are typically limited to $32 \times 32$ elements with 1024 channels \cite{verasonics_matrix_arrays} \cite{vermon_matrix_arrays}, and often require additional multiplexing \cite{matrix_array_mux}. In comparison, a $128 \times 128$ element \ac{tobe} array requires only 256 channels. Matrix probes with embedded microbeamformers such as Philips' xMatrix \cite{xmatrix_array} \cite{philips_xmatrix_page} can reduce the channel count but require complex in-probe electronics \cite{matrix_array_asic} that are difficult to scale to larger arrays. Sparse 2D arrays have also been proposed to reduce channel count \cite{sparse_spiral_array} \cite{costas_array} but suffer from low SNR and poor image quality \cite{sparse_array_review}.

\subsection{\acf{forces} Imaging} \label{sec:forces_background}

\subsubsection{The Multistatic Dataset and Hadamard Aperture Encoding}
The goal of \ac{forces} is to produce a "multistatic" dataset for a 1D linear array, a set of signals representing every combination of transmit and receive elements \cite{multistatic}. This dataset can be defined as an $N\times M$ matrix of signals: 
\begin{equation}
    \label{eq:multistatic}
    \sigmatm{S}(t) \in \mathbb{R}^{N \times M}
    \qquad
    \sigmatm{S}(t) =
    \begin{bmatrix}
        s_{1,1}(t) & \cdots & s_{1,M}(t) \\
        \vdots     & \ddots & \vdots \\
        s_{N,1}(t) & \cdots & s_{N,M}(t)
    \end{bmatrix}
\end{equation}
where $N$ and $M$ are the number of transmit and receive elements respectively, with $s_{ij}(t)$ representing the signal received on element $j$, transmitted by element $i$. 

In \ac{forces}, we receive on all channels along the lateral direction, like with standard \acp{rca} (Figure \ref{fig:tobe_array}B). This provides a unique signal for all $M$ receive elements, while the contributions of the $N$ transmitting elements are combined together. Standard \ac{sta} imaging separates these contributions by firing one element per event. After $N$ events, the full multistatic dataset has been recorded and is directly beamformed. This is illustrated in Figure \ref{fig:sequence_diagrams}A with an 8-element linear array. The remainder of this analysis focuses on how \ac{forces} separates these transmit contributions; for simplicity, we consider only a single receive channel ($M=1$).

\ac{forces} transmits on all lateral channels each event, with apodizations taken from a Hadamard matrix (Figure \ref{fig:sequence_diagrams}B) \cite{hadamard_review}. This is a square matrix consisting of only +1 and -1 elements, with the property that all rows and columns are mutually orthogonal. Hadamard matrices can be constructed with a recursive formula:
\begin{equation}
    \label{eq:hadamard-def}
    \begin{aligned}
        \bm{H}_2 = &\begin{pmatrix}
            +1,+1\\
            +1,-1
        \end{pmatrix} \\
        \bm{H}_{2^{a+b}} = \bm{H}_{2^a} &\otimes \bm{H}_{2^b},\ a,b\in \mathbb{N}
    \end{aligned}
\end{equation}
Note that this construction always yields a diagonally symmetric matrix with $\bm{H} = \bm{H}^T$. It is an open conjecture that a Hadamard matrix exists for all multiples of four, while the construction in equation \ref{eq:hadamard-def} is limited to powers of two. 

If a Hadamard apodization pattern is applied to the aperture for each event, the resulting dataset \sigmatt{G}(t) can be thought of as a linear combination of the multistatic dataset \sigmatt{S}(t) weighted by the Hadamard matrix:
{
\setlength{\jot}{1pt}
\begin{gather}
    \mathrm{g}_{e}(t) = \sum_{i=1}^{N} h^{(N)}_{ei} s_{i}(t) \qquad \sigmatm{G}(t) = \bm{H}_N\sigmatm{S}(t) \label{eq:forces-encode} \\ 
    \sigmatm{S}(t) = \bm{H}_N^{-1} \sigmatm{G}(t) \label{eq:forces-decode}
\end{gather}
}
Where for each event $e$, the received signal $\mathrm{g}_{e}(t)$ is a linear combination of the signals from each transmit element $s_{i}(t)$ weighted by the corresponding element of the Hadamard matrix $h^{(N)}_{ei}$. Thus, as Hadamard matrices are invertible, the original multistatic dataset can be recovered by applying the inverse of the Hadamard matrix ($\bm{H}_N^{-1}$) to the received dataset.

Figure \ref{fig:sequence_diagrams}B demonstrates the \ac{forces} sequence. The physical transmit events (left column) are biased according to the Hadamard matrix and include an elevational delay profile. After multiplying the eight received signals by the inverse matrix ($\bm{H}_8^{-1}$), a dataset with the same lateral transmit geometry as STA is recovered (right column). The dataset is beamformed in the same way as STA but produces a higher-quality image.
\\
\subsubsection{Delay and Sum Beamforming} \label{sec:das}
Once the multistatic dataset \sigmatt{S}(t) is recovered, it can be beamformed to produce an image. The most common method of beamforming is known as \acf{das}, which, for a given pixel, calculates the total transmit-pixel-receive path length for every transmit-receive pair, samples the signal at the corresponding time, and sums all contributions together to form the pixel value \cite{multistatic} \cite{das_review}.

The \ac{das} reconstruction on the multistatic dataset \sigmatt{S}(t) is defined for a single receive channel as:
\begin{equation}
\begin{gathered}
    \label{eq:base_das}
    \mathbb{D}\{\sigmatm{S}(t),r\} = \sum_{i=1}^{N}a(r) s_{i}(\tau_{i}(r)) \\
    \tau_{i}(r) = \frac{|d_{i}(r)| + |d_{Rx}(r)|}{c}
\end{gathered}
\end{equation}
Where $r$ is the location of a point in the imaged region, $N$ is the total number of transmitting elements indexed with $i$, and $a(r)$ is the dynamic receive apodization applied for point $r$. The delay function $\tau_{i}(r)$ takes the distance from transmitting \textbf{element} $i$ to point $r$ ($d_{i}(r)$) back to the receive element ($d_{Rx}(r)$) and converts it to time, where $c$ is the speed of sound in the medium. The form of $\tau_{i}(r)$ is dependent on the transmission focusing scheme. 

To beamform the \ac{forces} dataset, we simply apply the \ac{das} operator (\ref{eq:base_das}) to the decoded dataset (\ref{eq:forces-decode}):
\begin{equation}
\begin{gathered}
    \label{eq:das-forces}
    \mathbb{F}\{ \sigmatm{G}(t),r\} = \mathbb{D}\{\bm{H}_N^{-1} \sigmatm{G}(t),r\} \\
    \mathbb{F}\{ \sigmatm{G}(t),r\} = \sum_{i=1}^{N} \{\sum_{e=1}^{N} \hat{h}^{(N)}_{ie} g_{e}\}(\tau_{i}(r))
\end{gathered}
\end{equation}
where $\hat{h}^{(N)}_{ie}$ is entry $(i,e)$ in $\bm{H}_N^{-1}$ and $g_e$ is the encoded signal from \textbf{event} $e$. We refer to $\mathbb{F}\{ \sigmatm{G}(t),r\}$ as the "\ac{forces} reconstruction operator". \\

\begin{figure*}
    \centering
    \includegraphics[width=\linewidth]{figures/drawings/sequence_diagrams.pdf}
    \caption{Comparison between 8 transmit \ac{sta} (A), \ac{forces} (B), and \ac{readi} (C) sequences. The biasing for each \ac{forces} transmit event is taken from the rows of the $\bm{H}_8$ matrix at the bottom. \ac{readi} breaks the FORCES sequence into groups of 2, and partially decodes them using $\bm{H}_2^{-1}$. Each partially decoded group is beamformed into a READI low-resolution image, which are then summed to form the FORCES image. The \ac{forces} transmit events also include elevational delay focusing, which is annotated in the first event of B and represented by the arcing columns in the decoded events.}
    \label{fig:sequence_diagrams}
\end{figure*}

\subsubsection{Benefits and Challenges of FORCES} \label{sec:forces_benefits}

The primary benefit of \ac{forces} compared to other 2D STA sequences is the increased SNR from the Hadamard aperture encoding. By using all $N$ elements during each transmit event, the SNR improves by a factor of $\sqrt{N}$. \ac{tobe} arrays also enable elevational transmit focusing/steering. Since Hadamard apodization is implemented by applying DC biases to the lateral contacts, the transmitted signals can be routed to the elevational contacts with a fixed-delay focus. This delay profile (illustrated in the first event of Figure \ref{fig:sequence_diagrams}B) ensures that the maximum amount of energy is transmitted into the plane of imaging.  

The row-column symmetry of \ac{tobe} arrays allows the imaging plane to be rotated by 90 degrees without moving the array. This enables ”cross-plane” \ac{forces} sequences in which the transmit and receive directions swap between frames or transmission events, yielding two orthogonal high-resolution images. Additionally, the elevational focus can be "walked" across the array to form a 3D volume from a stack of \ac{forces} images. This "walking" is not achievable on standard RCAs, as their lack of bias control prevents the simultaneous implementation of lateral Hadamard transmit apodization with elevational transmit delay. Cross-plane and walking \ac{forces} have previously been compared with volumetric imaging schemes such as linear array mechanical scanning \cite{forces_ceroici_3d_comparison}, \ac{rca}-based plane and diverging wave sequences \cite{forces_comp_preprint}, and Explososcan on a fully-wired array \cite{forces_ceroici}. In all cases, \ac{forces} is shown to produce images with improved resolution, higher \ac{snr} and lower sidelobe levels. 

The large number of transmit events per \ac{forces} image, combined with events mixing during aperture decoding, can result in significant motion artifacts. A sparse transmit version of \ac{forces}, termed \ac{uforces}, has been proposed \cite{uforces}. However, the resulting images have a drastically reduced \ac{snr} and penetration compared with \ac{forces}. Another method to compensate for motion is desired that maintains \ac{forces}' high-quality structural images while also resolving motion over a large region. This is the motivation for \ac{readi} and \emcTwo.

\section{Methods} \label{sec:methods}
This section introduces the theory of \ac{readi} decoding and beamforming, which yield multiple low-resolution images. It then describes the process of estimating and compensating for the underlying motion, named \emcTwo. Finally, it details the implementation of the two algorithms on NVIDIA GPUs.

\subsection{\ac{readi} Decoding and Beamforming} \label{sec:readi-derivation}

\ac{readi} reduces the effect of motion blurring in Hadamard aperture-encoded sequences by beamforming subsets of the complete ensemble taken over smaller periods. This will be derived and tested in the context of \ac{forces} but can be applied to any Hadamard aperture-encoded sequence. 

Due to its recursive definition (equation \ref{eq:hadamard-def}), a rank $N$ Hadamard matrix can be expressed as a Kronecker product of rank $S$ and $Q$ Hadamard matrices, provided $S$ and $Q$ are both powers of two \cite{hadamard_review}. 

\begin{gather} 
\begin{aligned} \label{eq:hadamard-recurse}
	\bm{H}_N &= \bm{H}_S \otimes \bm{H}_Q \\ 
    \bm{H}_N^{-1}&=\bm{H}_S^{-1} \otimes \bm{H}_Q^{-1} \\
    N&=QS  
\end{aligned}
\end{gather}
where $\bm{H}_{S}$ and $\bm{H}_{Q}$ are smaller Hadamard matrices that form $\bm{H}_{N}$ while $\bm{H}_{N}^{-1}$, $\bm{H}_{S}^{-1}$, and $\bm{H}_{Q}^{-1}$ are their inverses. 

According to the definition of the Kronecker product, any individual element of $\bm{H}_{N}$ (denoted as $h^{(N)}_{e,i}$) can be expressed as the scalar product of an element from each submatrix:
\begin{equation} \label{eq:kronecker_index}
   h^{(N)}_{e,i} = h^{(N)}_{(q+Q(s-1)),(v+Q(l-1))} = h^{(S)}_{s,l} h^{(Q)}_{q,v}
\end{equation}
where $h^{(S)}_{s,l}$ and $h^{(Q)}_{q,v}$ are individual elements of $\bm{H}_{S}$ and $\bm{H}_{Q}$ respectively.

To apply this definition to \ac{forces}, we divide the $N$ transmit events sequentially into $S$ groups of $Q$ transmit events. The transmit \textbf{event} $e$ is thus reindexed by its group number $s$ and its index within the group $q$. Similarly, each transmit \textbf{element} $i$ reindexed by its group number $l$ and its index within the group $v$:

\begin{equation}
\begin{aligned} \label{eq:reindex}
    e &= q + Q(s-1) \\
	i &= v + Q(l-1)
\end{aligned}
\end{equation}

With the new event indexing, we reshape the $1 \times N$ \ac{forces} dataset \sigmatt{G}(t) into a $Q \times S$ matrix \sigmatt{G'}(t), where each column contains the data from one group of $Q$ transmit events:
\begin{equation}
\begin{array}{c}
\sigmatm{G}(t) \in \mathbb{R}^{N \times 1} \to \sigmatm{G'}(t) \in \mathbb{R}^{Q \times S} \\
\sigmatm{g'}_{s}(t) := \mathrm{col}_{s} \sigmatm{G'}(t) \  (s = 1, \ldots, S)
\label{eq:grouped-data}
\end{array}
\end{equation}
where $\sigmatm{g'}_{s}(t)$ is a column vector containing the $Q$ signals from the $s$'th group of transmit events.

Applying equations \ref{eq:hadamard-recurse} - \ref{eq:grouped-data} to the \ac{forces} reconstruction operator (\ref{eq:das-forces}) yields:
\begin{equation}
\mathbb{F}\{  \sigmatm{G'}(t),r\} =  \sum_{s}^S \tikzmarknode[outer ysep=5pt]{f_left}{[}\sum_{l}^{S} \hat{h}^{(S)}_{l,s}\mathbb{D'}_{l} \{ \bm{H}_Q^{-1} \sigmatm{g'}_{s}(t), r \} \tikzmarknode[outer ysep=5pt]{f_right}{]}
\label{eq:forces-das-recursed} 
\end{equation}
\vspace{1em}
\tikzset{annotate equations/arrow/.style={-}}
\annotatetwo[yshift=-1em]{below, label below}{f_left}{f_right}{\textit{Q} event low-resolution image}

Here $\hat{h}^{(S)}_{l,s}$ is an element of $\bm{H}_S^{-1}$ and $\mathbb{D'}_{l}$ is a modification of the \ac{das} operator to account for the grouping of transmit elements. The derivation of this equation is provided in Appendix I.

This new form of \ac{forces} reconstruction produces a low-resolution image from each group of transmit events, then sums them to form the high-resolution \ac{forces} image. We have named the term in square brackets that generates low-resolution images the \textbf{READI Reconstruction Operator}.

\begin{subequations} \label{eq:readi-full}
\begin{equation} \label{eq:readi-compound}
    \mathbb{F}\{ \sigmatm{G'}(t),r\} = \sum_{s}^S \mathbb{R}\{  \sigmatm{g'}_s(t),r\}
\end{equation}
\vspace{1em}
\begin{equation} \label{eq:readi-lr}
\mathbb{R}\{  \sigmatm{g'}_s(t),r\} = \tikzmarknode[outer ysep=5pt]{low_left}{\sum_{l}^{S}} \hat{h}^{(S)}_{l,s} \tikzmarknode[outer ysep=5pt]{sub_left}{\mathbb{D'}}_{l} \tikzmarknode[outer ysep=5pt]{dec_left}{\{} \bm{H}_Q^{-1} \sigmatm{g'}_{s}(t)\tikzmarknode[outer ysep=5pt]{dec_right}{,} r \tikzmarknode[outer ysep=5pt]{sub_right}{\}}
\end{equation}
\vspace{2em}
\tikzset{annotate equations/arrow/.style={-}}
\annotatetwo[yshift=-1em]{below, label below}{dec_left}{dec_right}{1. Partial Decode}
\annotatetwo[yshift=1em]{above, label above}{sub_left}{sub_right}{2. Sub-Image Beamform}
\annotatetwo[yshift=-2em]{below, label below}{low_left}{sub_right}{3. READI Low-Resolution Image}
\end{subequations}

We interpret the \ac{readi} reconstruction operator as consisting of three stages:
\begin{enumerate}
    \item Partially decode of the $s$'th group of $Q$ transmit events with $\bm{H}_Q^{-1}$.
    \item Beamform the partially decoded event group into $S$ sub-images, one for each element group $l$.
    \item Sum these sub-images into a \ac{readi} low-resolution image, weighted with the $s$'th row of $\bm{H}_S^{-1}$.
\end{enumerate}

Figure \ref{fig:sequence_diagrams}C highlights the \ac{readi} beamforming process and contrasts it with \ac{sta} and \ac{forces} (Figures \ref{fig:sequence_diagrams}A-B). The same \ac{forces} transmit sequence is broken into $S=4$ groups of $Q=2$ transmit events. Each group is partially decoded with $\bm{H}_2^{-1}$ into a set of sparse subapertures (stage 1), then beamformed into a \ac{readi} low-resolution image using equation \ref{eq:readi-lr} (stages 2-3). The four \ac{readi} images are then summed to obtain the final image, which is identical to the output of the standard \ac{forces} method. Note that each partially decoded group has a different polarity pattern across its active transmit elements. These patterns correspond to the rows of $\bm{H}_4$, and are the source of each group's unique READI artifacts, discussed later. 

Critically, each \ac{readi} low-resolution image is only formed by one group \sigmatt{g'}\textsubscript{s}(t) of $Q$ transmit events, rather than the complete set of $N$ events. Therefore, any motion artifacts present in the \ac{forces} image will be reduced by a factor of $N/Q$. This allows a single transmit sequence to produce a high-resolution image and a set of low-resolution images with good motion fidelity. As the READI operator is a linear reorganization of the FORCES operator, the compounded READI images should be mathematically identical to the FORCES image, within floating-point precision. Appendix II analyzes equation \ref{eq:readi-full} to demonstrate how this perfect reconstruction is achieved.

\subsection{Estimated Motion Compensated Compounding (\ac{emc2})}
Any feature that changes position between READI images will appear blurred in the FORCES image after compounding. If, instead, the $S$ \ac{readi} images are warped to align with one another before compounding, they can be summed to form a high-quality FORCES image without motion artifacts. We refer to this new process as\acreset{emc2} \textbf{\ac{emc2}}.

\begin{figure}
    \centering
    \includegraphics[width=\linewidth]{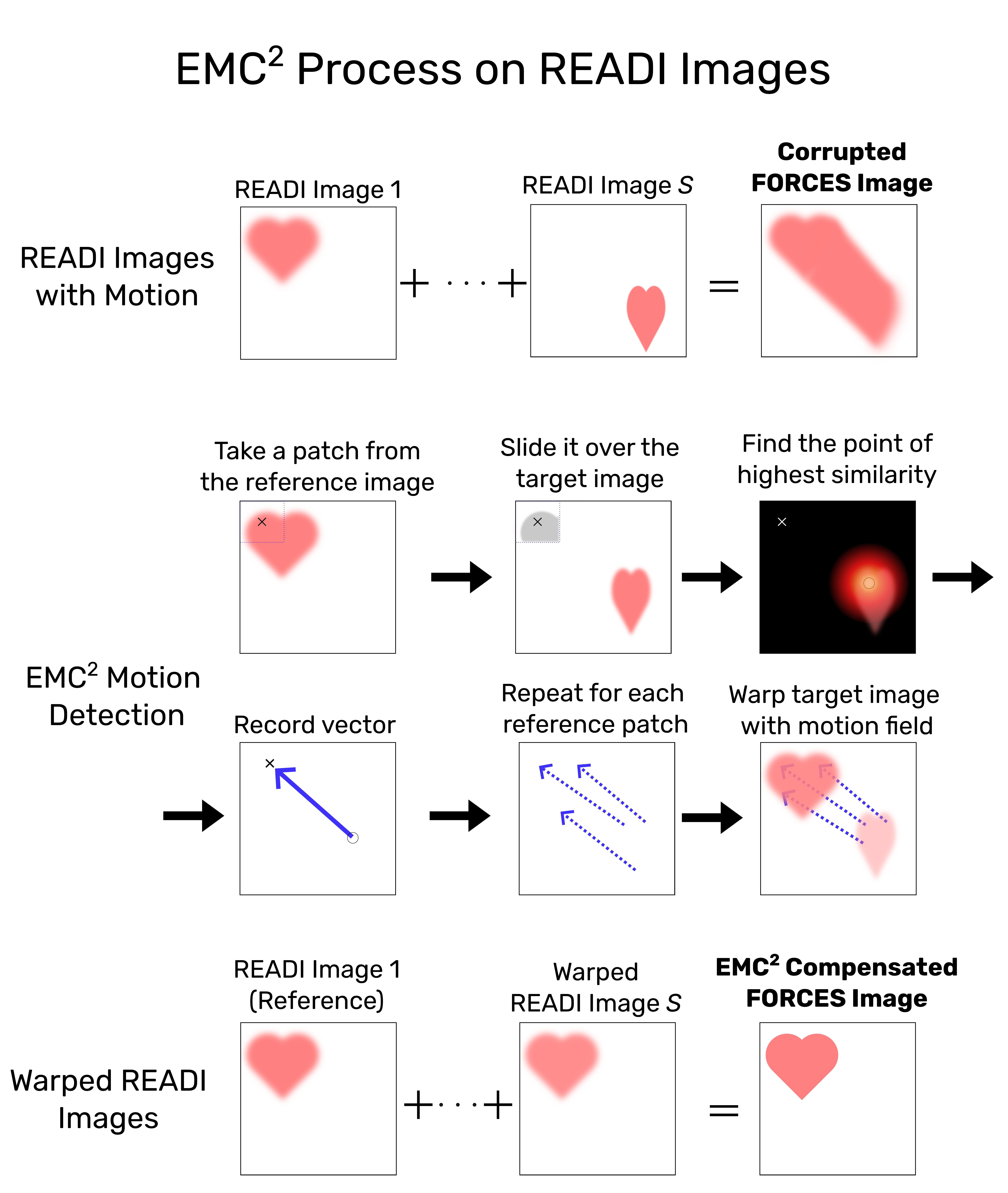}
    \caption{Example \emcTwo\  motion compensation process. The first \ac{readi} image is taken as the reference and not warped.}
    \label{fig:emc2_example}
\end{figure}

\ac{emc2} consists of two steps: motion detection and warping. The \ac{readi} images are compared to one another using a motion detection algorithm, yielding a motion vector for each point in each image. The images are then warped using these motion vectors to align them spatially before being compounded into the \ac{forces} image. This yields an image that is both high-resolution and robust to motion blur. The basic process is illustrated in Figure \ref{fig:emc2_example}. \emcTwo\ can use any motion-detection algorithm that operates on a sequence of beamformed images. For this work, we select \acreset{ncc}\ac{ncc} based block matching \cite{ncc_block_matching} due to its robustness to local brightness and contrast changes \cite{block_matching_comparison}, but many other options are available and will be explored in future work.

Table \ref{tb:method_comp} summarizes the benefits and drawbacks of the methods discussed in the previous sections.
\begin{table}
    \centering
    \begin{tabular}{||m{4em}|m{10em}|m{10em}||} 
         \hline
         Method & Pros & Cons \\
         \hline \hline
            Base STA & High Resolution, Fast, Simple Implementation & Low SNR, Bad Penetration, Sensitive to Motion \\
         \hline
            FORCES & High Resolution and SNR, Good Penetration, Fast & Sensitive to Motion \\
         \hline
            uFORCES & Ultrafast, Low Motion Sensitivity & Low Resolution and SNR, Bad Penetration\\
         \hline
            READI with EMC$^2$ & High Resolution and SNR, Good Penetration, Fast, Robust Motion Compensation & Complex Beamforming\\
         \hline
    \end{tabular} 
    \vspace{0.5em}
    \caption{Synthetic Transmit Aperture Method Comparison}
    \label{tb:method_comp}
\end{table}

\subsection{Implementation}
All beamforming and motion compensation is performed on NVIDIA GPUs using the CUDA\cite{nvidia_cuda} platform. The application utilizes custom CUDA kernels for beamforming and motion estimation. Additionally, the cuBLAS \cite{nvidia_cublas}, cuFFT \cite{nvidia_cufft}, and \ac{npp} \cite{nvidia_npp} libraries provide well-optimized kernels for matrix multiplication, Fourier transforms, and image comparison, respectively. Array details are provided later with the results.\\

\subsubsection{Acquisition Rate}
All data was collected on a Verasonics Vantage 256 system with custom DC-biasing electronics \cite{forces_comp_preprint} \cite{tobe_bias_electronics} and transferred to the CUDA application for processing. The \ac{prf} of \ac{forces} acquisitions is a combination of round trip-time to the maximum depth ($2t_{z_{max}}$), and the time it takes for the biasing electronics to switch states ($t_{bias}$), the frame rate is determined by that and the number of transmissions per frame ($N$):
\begin{equation} 
        f_{forces} = \frac{f_{prf}}{N} = \frac{1}{N(2t_{z_{max}} + t_{bias})} 
\end{equation}

Applying \ac{readi} multiplies this frame rate by the number of groups $S$. Currently implemented electronics have a maximum switching time of $250\ \mu s$ \cite{tobe_bias_electronics}, limiting the maximum \ac{prf} to \textasciitilde 3 kHz and FORCES frame rate to \textasciitilde 24 FPS for a 128x128 element array at an imaging depth of \textasciitilde 60 mm. Breaking up the sequence into groups of 8 with READI would increase this limit to \textasciitilde 375 FPS. A new design of these electronics under development aims to reduce this delay by an order of magnitude. Experiments in this work use PRFs of 0.5 - 3.5 KHz at various imaging depths.\\

\subsubsection{\ac{readi} Beamformer}
The \ac{readi} beamforming process performs the following steps:
\renewcommand{\labelenumi}{\Alph{enumi})}
\renewcommand{\labelenumii}{\roman{enumii})}
\begin{enumerate}
\item Select $S$ and $Q$ such that $N=SQ$, break the data into $S$ sequential groups of $Q$ transmissions
\item Partially decode each group ($\sigmatm{g'}_s(t)$) with $\bm{H}_Q^{-1}$ using \texttt{cublasSgemm} matrix multiplication
\item Hilbert transform the partially decoded data to obtain the RF analytic signal \cite{hilbert_analytic}
\begin{enumerate}
    \item Take the real to complex Fourier transform of the signal (cuFFT)
    \item Zero all negative frequency bins and double all positive frequency bins (custom kernel)
    \item Take the complex to complex inverse Fourier transform of the modified spectrum (cuFFT)
\end{enumerate}
\item Apply the \ac{readi} reconstruction operator (eq. \ref{eq:readi-lr}) to each partially decoded group (custom kernels):
\begin{enumerate}
    \item Beamform the group $\sigmatm{g'}_s(t)$ into $S$ sub-images, together containing all in-focus terms plus extra cross terms
    \item Create a sum of these sub-images weighted by the $s$'th row of $\bm{H}_S^{-1}$ to form the low-resolution \ac{readi} image
\end{enumerate}
\item Sum the low-resolution \ac{readi} images from all groups to form the final \ac{forces} image
\end{enumerate}

See Appendix II for an explanation of step "D". Both the low-resolution \ac{readi} images and the full \ac{forces} image are stored in their complex RF form for further processing.

Optionally, each pixel is weighted by its coherence factor during beamforming to improve \ac{snr}. This is a method for estimating the coherence between each sample $S_k$ forming the pixel \cite{coherence_factor}. It is defined as:
\begin{equation*}
    \mathrm{CF} = \frac{|\sum S_{k}|^2}{\sum|S_{k}|^2}
\end{equation*}
As coherence factor is a non-linear operation, applying it to the \ac{readi} low-resolution images during beamforming yields a slightly different result than applying it directly during \ac{forces} beamforming. Accordingly, it is disabled when appropriate for comparison. \\ 

\begin{figure*}[t]
    \centering
    \includegraphics[width=\linewidth]{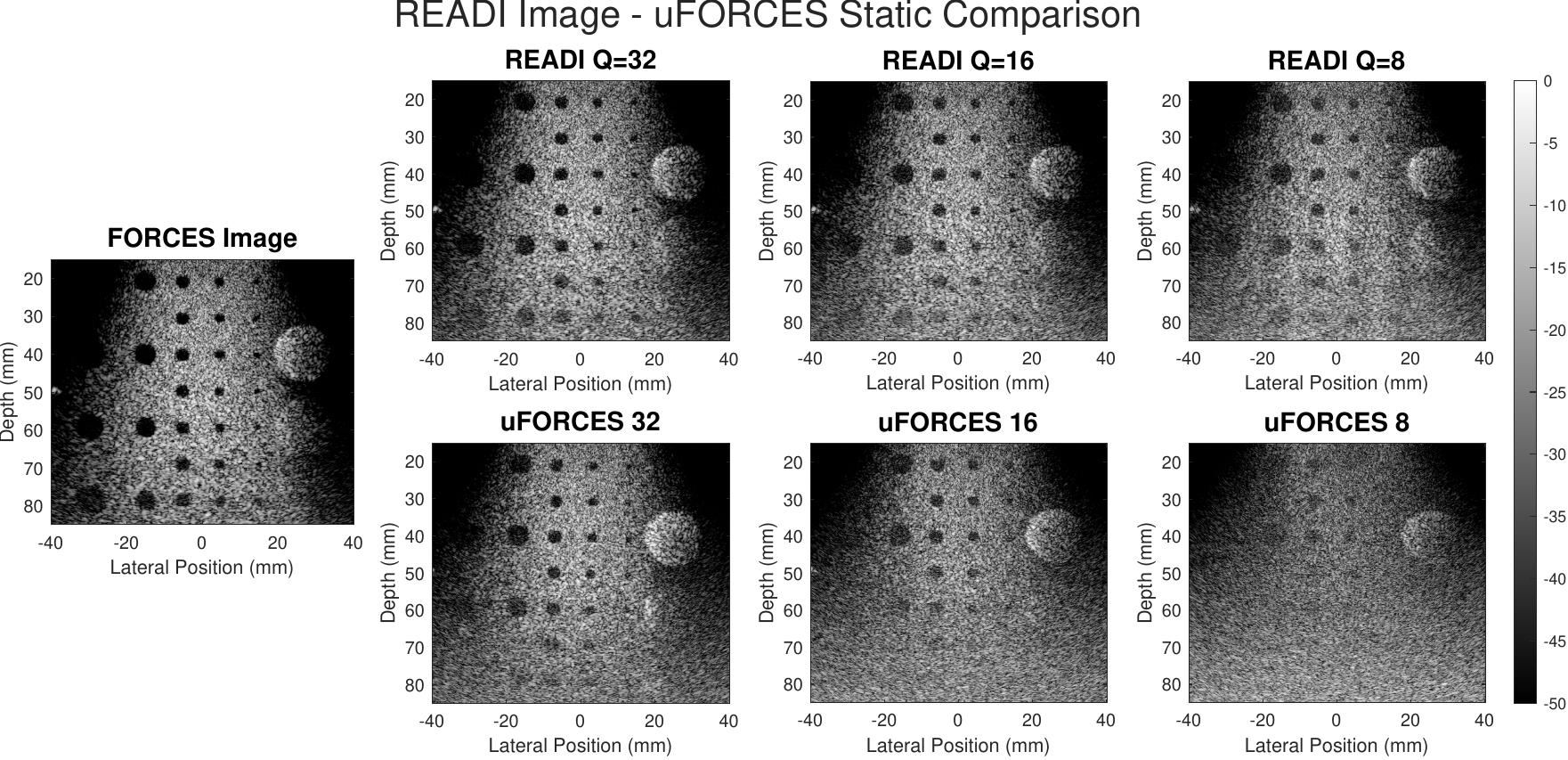}
    \caption{Comparison between READI low-resolution images and uFORCES images with the same transmit count. Left to right: 128 transmit FORCES image, 32 transmit comparison, 16 transmit comparison, 8 transmit comparison.}
    \label{fig:static_readi_uforces_comp}
\end{figure*}

\begin{figure}
    \centering
    \includegraphics[width=\linewidth]{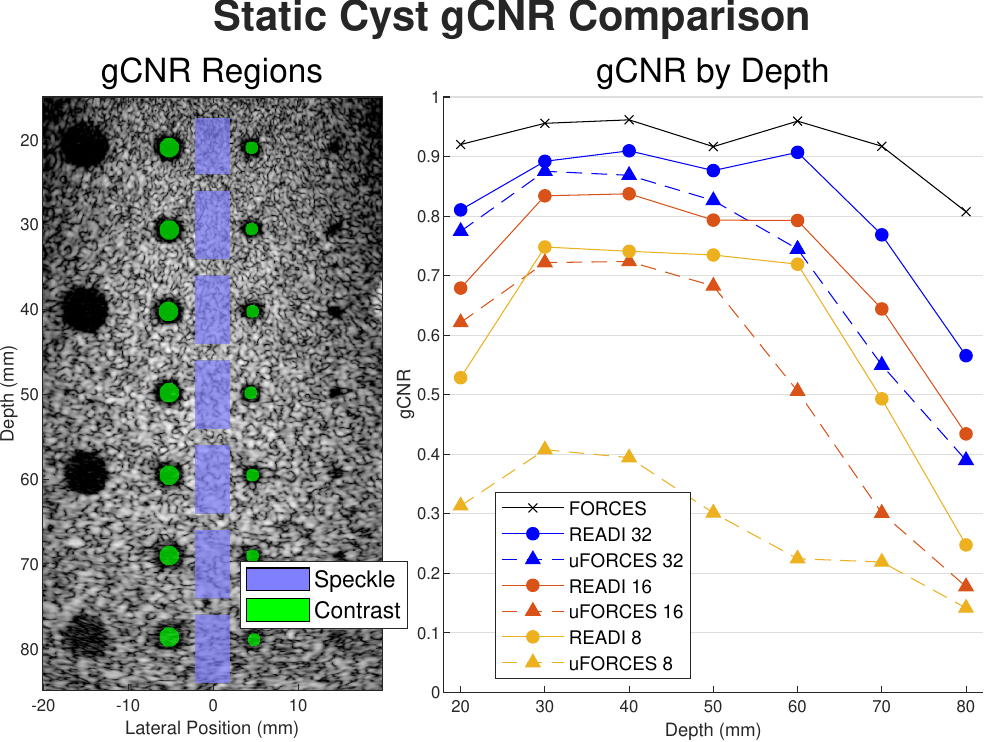}
    \caption{gCNR measurements for the static cyst phantom comparison, regions of interest for each depth are marked on the left.}
    \label{fig:static_gcnr}
\end{figure}

\subsubsection{Motion Compensation} \label{sec:motion_compensation}
Our \emcTwo\ implementation of block matching utilizes a CUDA-based version of \ac{ncc} \cite{ncc_block_matching} from the \ac{npp} image processing library. One of the $S$ images is selected as the reference, and all other images (targets) are compared with it using \ac{ncc}. Motion is estimated at points with 4 to 16 pixel spacing to reduce computational cost; the output motion field is then interpolated up to the image's resolution before warping. 

As shown in Figure \ref{fig:emc2_example}, at each location in the motion grid, a patch is extracted from the reference image centred on the point. This reference patch is slid over a larger region of the target image, and the correlation is computed at each position. This outputs a similarity map of values in the range [-1,1], with 1 indicating a perfect match. The peak value in this map corresponds to the location in the target image that best matches the reference patch. The motion value at this location is the vector from the peak to the center of the reference patch.

The correlation map for each reference patch is calculated with the \ac{npp} library's \texttt{nppiCrossCorrValid\_NormLevel} method. The peak of each similarity map is identified using a custom CUDA kernel; the sub-pixel peak location is obtained by fitting a 2D paraboloid to a 3x3 region centred on the peak value \cite{curve_fit}.

While \ac{ncc} is resilient to changes in local brightness and contrast \cite{block_matching_comparison}, it is still susceptible to noise corrupting the comparisons. To reject spurious motion vectors, three criteria are applied to the correlation surface around the detected peak:
\begin{itemize}
    \item The peak value must exceed a predefined absolute threshold (reject guessing in noisy regions with weak correlation).
    \item The peak must also be a set percentage above the non-motion value (reject motion if the match is only slightly stronger than the non-motion).
    \item The curvature of the fitted paraboloid must exceed a minimum (reject flat surfaces where exact motion is hard to guess).
\end{itemize}
After this process, each motion field is interpolated up to the image resolution and used to warp its source image. Summing these warped images produces a full \ac{forces} image with drastically reduced motion artifacts.

Table \ref{tb:ncc} lists the parameters used for motion detection and their ranges. These parameters were manually tuned for each experiment to maximize contrast and image sharpness in motion-corrupted regions, guided by comparisons with the static cases and subjective visual quality. These metrics are described in more detail in the results section.

\renewcommand{\arraystretch}{1.5}
\begin{table}
    \centering
    \begin{tabular}{||c|c||} 
         \hline 
         Motion Grid Size & 4-16 px \\ 
         \hline 
         Reference Patch Size & 16-64 px \\
         \hline
         Search Patch Size & Ref. size + 8-64 px \\
         \hline
         Absolute Peak Threshold & 0.01 - 0.5 \\
         \hline
         Relative Peak Threshold & 101\% - 120\% \\
         \hline
         Minimum Curvature & 0.005 - 0.1 \\
         \hline
    \end{tabular} 
    \vspace{0.5em}
    \caption{NCC Motion Detection Parameters}
    \label{tb:ncc}
\end{table}

\section{Results} \label{sec:Results}

All \ac{forces} data was collected using a 128x128 element lambda pitch \ac{tobe} array produced by CliniSonix Inc. on a Verasonics Vantage 256 system with CliniSonix DC biasing electronics \cite{forces_comp_preprint}. Figure \ref{fig:tobe_array}D provides a block diagram of this setup.
Table \ref{tb:experiment_params} lists the imaging parameters. These are valid for all experiments unless otherwise noted in their individual sections.

Three primary metrics are used to evaluate the results: \acs{gcnr}, speckle similarity, and image sharpness.
\ac{gcnr} \cite{gcnr} measures the overlap between the probability density function of pixels within a contrast region and that of pixels in a background speckle region. A \ac{gcnr} of 1 indicates no overlap between these distributions and perfect contrast, while a \ac{gcnr} of 0 indicates complete overlap and no contrast. 
This method is also used to compare the speckles of the corrupted and \emcTwo-compensated images; here, the scale is flipped so that 1 indicates complete overlap of the probability densities (and thus recovery of the original speckle distribution). We refer to this comparison as "Speckle Similarity" in the results. 
Finally, image sharpness is measured from the spatial Fourier domain \cite{fourier_sharpness}. It is calculated as the ratio of significant frequency components to the total number of frequency components. Sharp images will contain contributions from a wide range of frequencies, whereas blurry images will contain fewer significant frequencies.
Speckle similarity is calculated on the linear beamformed images; all other methods are calculated after log-compression.

\begin{table}
    \centering
    \begin{tabular}{||c|c||} 
        \hline
        FORCES Transmit Count & 128 \\
        \hline
        Transmit Frequency & 4.3 MHz \\
        \hline
        Element Pitch & $\lambda$ \\
        \hline
        Transmit Waveform & 2-cycle sinusoid \\
        \hline
        Pulse Repetition Frequency (PRF) & 1 kHz \\
        \hline
        Dynamic Rx Apodization F\# & 1 \\
        \hline
        Coherence Factor Weighting & Enabled \\
        \hline
        Pixel Size & $0.05 \times 0.05$ mm \\
        \hline
    \end{tabular} 
    \vspace{0.5em}
    \caption{Imaging Parameters}
    \label{tb:experiment_params}
\end{table}
\vspace{0.5em}

\begin{figure*}[t]
    \centering
    \includegraphics[width=\linewidth]{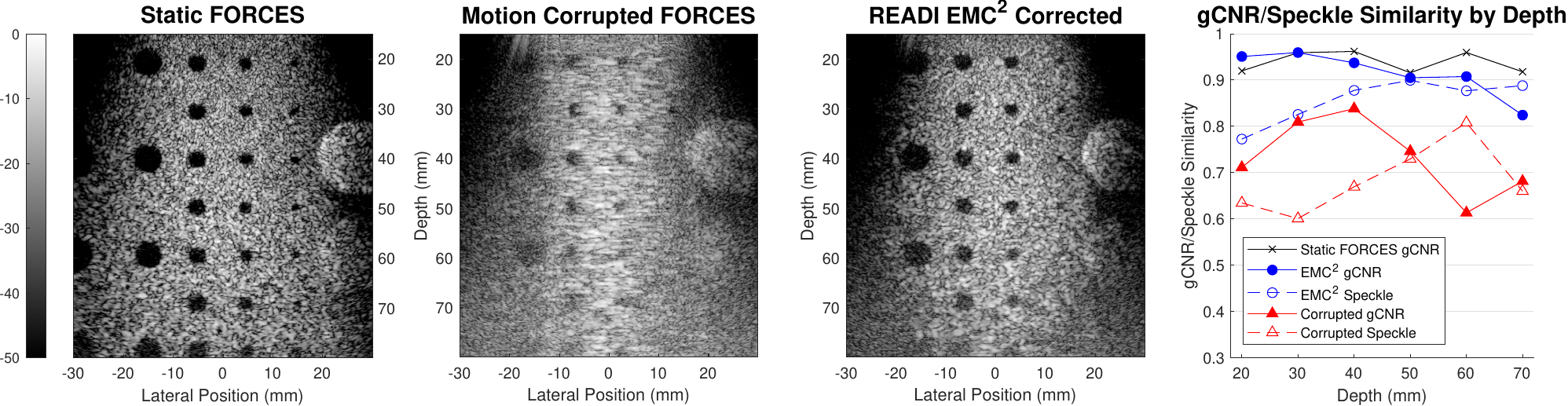}
    \caption{FORCES image before (left) and after (center) \ac{emc2} motion compensation. Right: gCNR measurements (solid lines) and speckle similarity (dashed lines) at each depth. Speckle similarity is calculated by comparison with the static case.}
    \label{fig:probe_motion}
\end{figure*}

\begin{figure*}
    \centering
    \includegraphics[width=\linewidth]{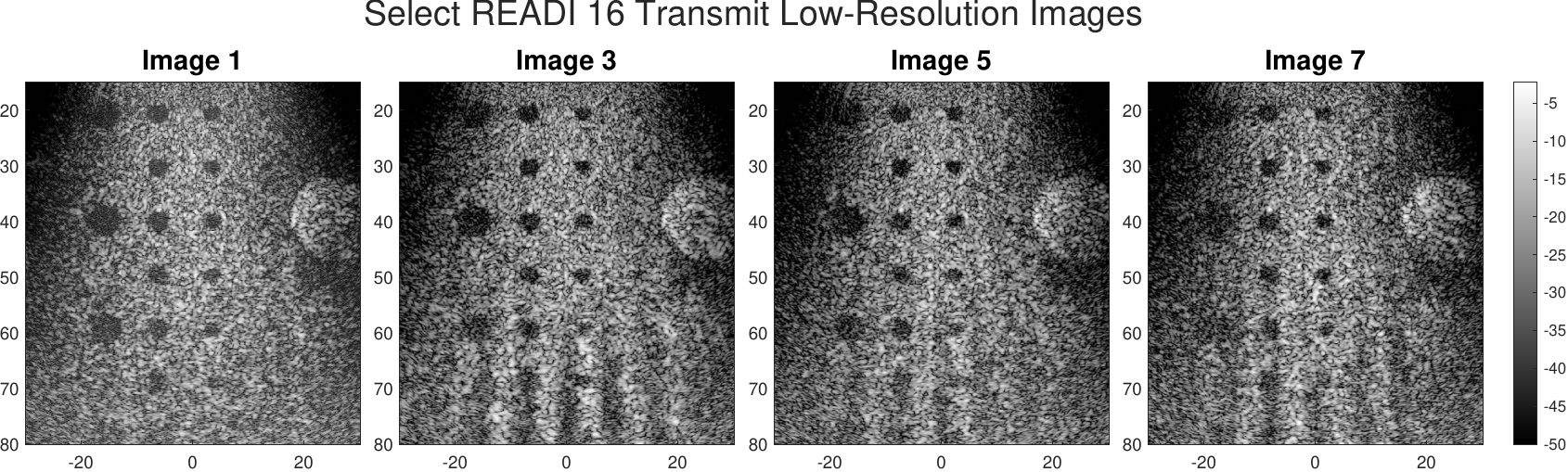}
    \caption{Select READI low-resolution images from the probe motion experiment. All images are formed from groups of 16 transmissions.}
    \label{fig:cyst_low_res}
\end{figure*}

\subsection{Field II Simulations} \label{sec:psf_simulation}

Multiple simulations were performed with Field II \cite{field_ii_1} \cite{field_ii_2} to validate the \ac{readi} beamforming process and to explore the effects of motion and resolution, their results are collected in Appendix III-A and III-B. First, the READI beamformer was validated with a static 128 transmit FORCES point-spread function simulation. At all tested group sizes ($Q=32,16,8$), the compounded READI image was identical to the FORCES image within floating-point precision, confirming the mathematical equivalence of the two operators discussed in section \ref{sec:readi-derivation}. Resolution was measured by simulating point spread function at various depths. The \ac{readi} images maintain the lateral resolution of \ac{forces} even with few transmits per group, but do show some decrease in axial resolution and sidelobe height as group size decreases. These results are included in Appendix III-Ain the supplementary materials.

Appendix III-B provides simulations of the effects of motion on \ac{readi} images, demonstrating that small READI group sizes can significantly improve resolution and enable motion estimation even when the \ac{forces} image is completely corrupted. Simulation parameters are also provided with their respective results. All simulation code is available upon request.
Coherence factor weighting was disabled for these noiseless simulations to preserve linearity, but is enabled for all subsequent results.

\subsection{Static Comparison} \label{sec:static_comparison}
For the first real-world experiment, an anechoic cyst phantom (\textit{ATS 539}, CIRS Inc., Norfolk, VA, US) was imaged with \ac{forces} and beamformed again with \ac{readi} using group sizes of 32, 16, and 8. The quality of the \ac{readi} low-resolution images was compared with \ac{uforces} images with the same transmit count. \ac{uforces} is a sparse transmission version of \ac{forces} that uses $Q < N$ Hadamard encoded transmissions \cite{uforces}. The decoded \ac{uforces} dataset contains $Q-1$ signals, each representing a single transmitting element. The leftover signal contains the contributions of all other $N-Q+1$ elements and is discarded. 
Figure \ref{fig:static_readi_uforces_comp} compares the first \ac{readi} image at each group size to a \ac{uforces} image formed with the same transmit count. Both image sets degrade as the transmit count decreases, but at each transmit count, the \ac{readi} image exhibits a visibly higher \ac{snr} than the \ac{uforces} image.

To quantify this quality difference, the \ac{gcnr} was computed for the middle two columns of cysts at each depth. The results for the seven images are plotted in Figure \ref{fig:static_gcnr} with the regions of interest marked. The \ac{readi} images consistently outperform \ac{uforces} at all depths and transmit counts, with the gap increasing as transmit count decreases.

\begin{figure*}[t]
    \centering
    \includegraphics[width=\linewidth]{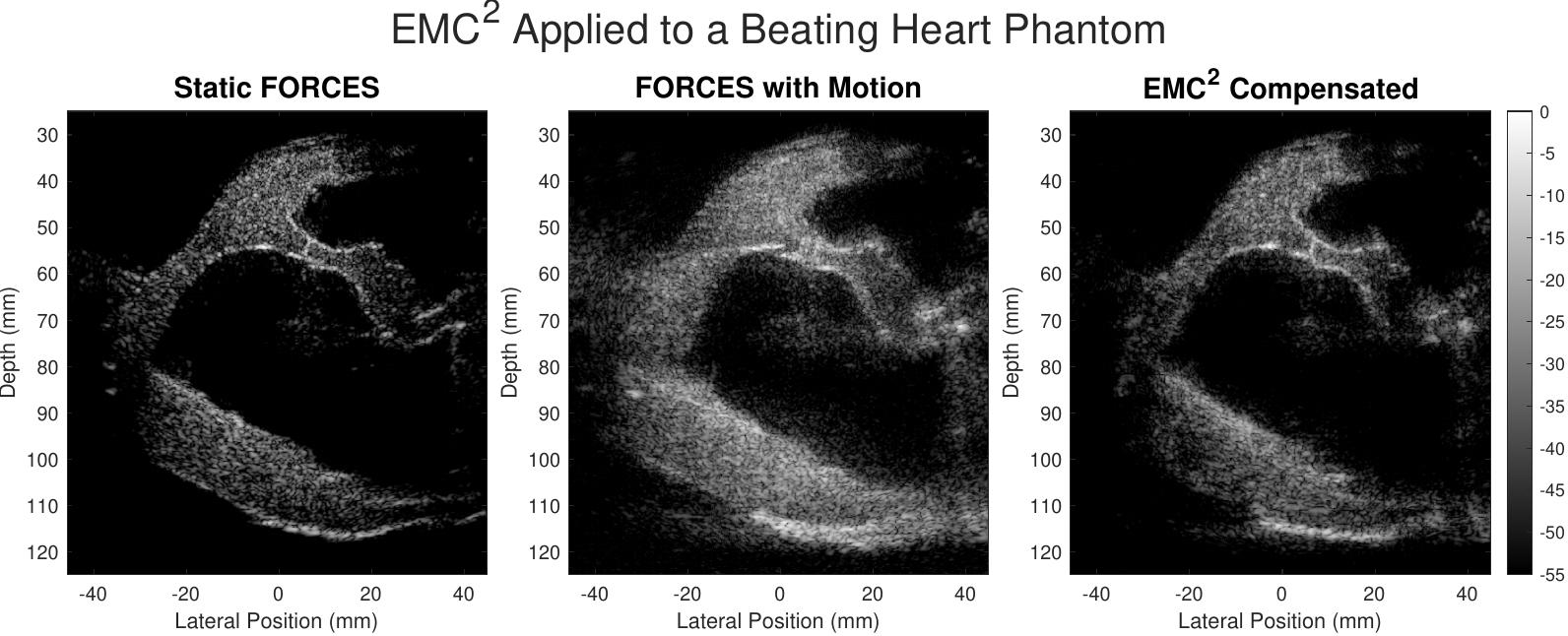}
    \caption{\ac{emc2} compensated image of a beating heart phantom (right) compared with the motion corrupted image (center) and a FORCES image of the static heart (left).}
    \label{fig:heart_comp}
\end{figure*}

\subsection{Probe Motion}
Next, the cyst phantom was imaged while the operator manually moved the probe laterally at an approximate speed of 18.5 mm/s. This resulted in a significantly corrupted \ac{forces} image where the cysts are barely visible. The data was then split into eight groups of sixteen transmissions ($ Q=16$, $ S=8$) and beamformed using \ac{readi}; \emcTwo\ was then applied to the eight low-resolution images. The \ac{emc2} compensated image shows a significant reduction in motion artifacts, with the cysts and background speckle clearly visible (Figure \ref{fig:probe_motion}). Figure \ref{fig:cyst_low_res} shows select \ac{readi} low-resolution images from this experiment. Each group has a unique brightness pattern introduced by READI decoding that appears as vertical striping. These READI artifacts worsen with depth and are more severe for small group sizes. 

The gCNR was measured for each image at the same locations marked in Figure \ref{fig:static_gcnr}, with results presented in Figure \ref{fig:probe_motion}. At each depth, the speckle of the corrupted and \emcTwo\ images were also compared with the static case; their Speckle Similarity scores are plotted as dashed lines. The sharpnesses of the three images were also measured between -20 and 20 mm, with results in Table \ref{tb:cyst_sharpness}. The \ac{emc2} compensated image successfully recovers the same \ac{gcnr} as the static image for all cysts clearly visible in the low-resolution images. Beyond a depth of 70mm, the cysts have been lost in the noise and \ac{readi} artifacts and are unrecoverable. The \emcTwo\ recovered speckle also shows improved similarity with the static case at each depth. Averaging the estimated motion between each of the eight \ac{readi} images yields an estimated average lateral speed of $17.6 \pm 1.3$ mm/s, which agrees with the expected value of 18.5 mm/s within error.

\begin{table}[h]
    \centering
    \begin{tabular}{||l|l|l||} 
         \hline
         Image & Sharpness & \% Error \\
         \hline \hline
            Static FORCES & 0.106 & N/A \\
         \hline
            Blurred FORCES & 0.045 & 58\% \\
         \hline
            \emcTwo\ Compensated & 0.069 & 35\% \\
         \hline
    \end{tabular} 
    \vspace{0.5em}
    \caption{Cyst Sharpness Comparison}
    \label{tb:cyst_sharpness}
\end{table}

This demonstrates that \ac{emc2} can recover a B-mode of similar quality to a static \ac{forces} image in the presence of bulk motion. Comparing the \ac{emc2} compensated cyst image with the static \ac{forces} image, the leftmost column of large cysts is outside the compensated image's field-of-view. While \ac{emc2} successfully resolves structures within the field-of-view, it does not return it to the width of the static case. Reducing the number of transmissions per group to $Q=8$ and adjusting the receive apodization did not recover the lost information. This indicates that \ac{readi} can easily recover information that remains within the field-of-view across the entire ensemble, but significant bulk motion can lead to information loss at the image edges.

\begin{figure*}[t]
    \centering
    \includegraphics[width=\linewidth]{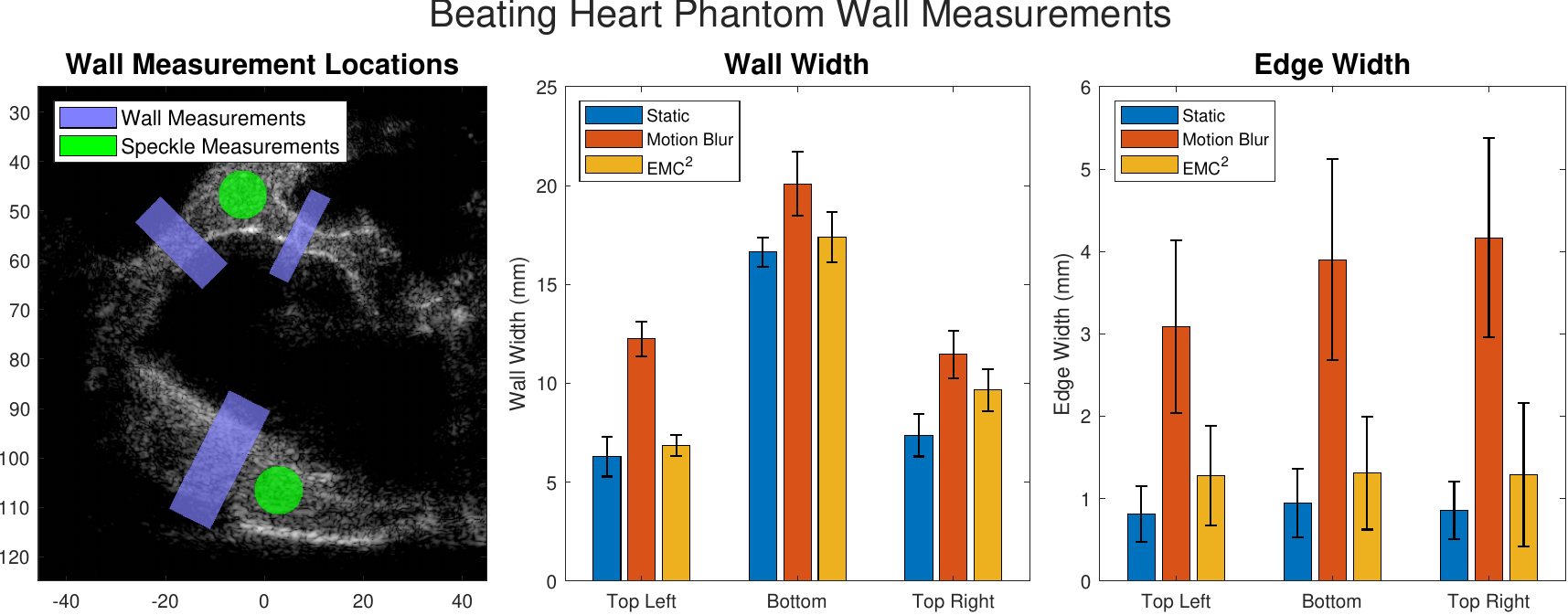}
    \caption{Wall width and edge thickness measurements for the marked regions on the heart phantom images in Figure \ref{fig:heart_comp}. Error bars indicate the standard deviation of these measurements across the marked regions.}
    \label{fig:heart_measurements}
\end{figure*}
\begin{figure}[h]
    \centering
    \includegraphics[width=\linewidth]{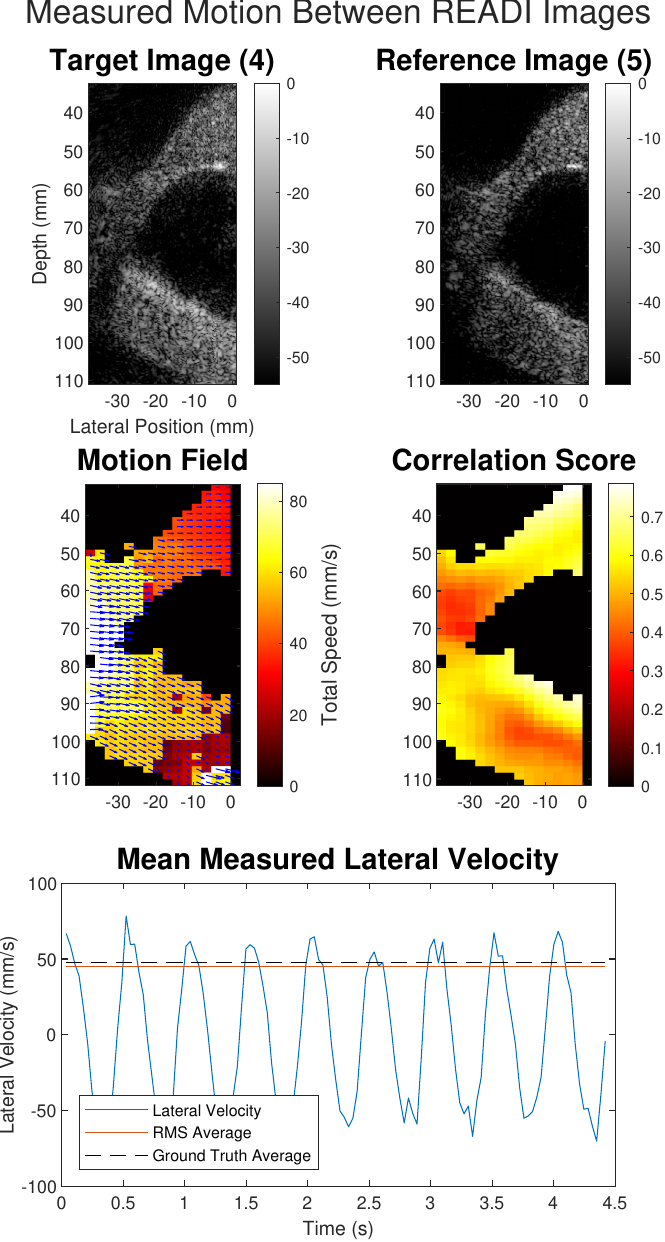}
    \caption{Top: Measured motion between the fourth and fifth READI images from the heart phantom experiment, taken from the left wall of the heart. The colour of the third image indicates the total speed, while the fourth indicates the correlation score of the estimated vector. Bottom: Average lateral velocity of the phantom on the left wall, across 16 FORCES frames (128 READI images).}
    \label{fig:motion_sample}
\end{figure}

\subsection{Beating Heart Phantom}

For a more realistic test case, a heart phantom (\textit{DHP01}, Shelley Medical Imaging Technologies, Toronto, ON, Canada) in a water bath was imaged as a servo simulated a 120 BPM heartbeat. This experiment was performed at 5 MHz with a \ac{prf} of 0.5 kHz. The servo arm is attached to the bottom of the heart (at x = -40 mm in the images) and is actuated to simulate the beating motion. While this phantom lacks true three-dimensional heart motion, it serves as a useful test case for \ac{readi}'s ability to resolve general tissue motion. The servo takes 0.3 seconds to complete a compression stroke and compresses the heart linearly by 14.3 mm. Thus, we can estimate the average lateral speed of the heart near the connection point (on the left side of the image) to be approximately $\frac{14.3}{0.3} = 47.7$ mm/s. During continuous motion, the phantom has a BPM of 120, leaving 0.25 seconds for each stroke (compression and retraction). Again \ac{readi} was applied using eight groups of sixteen ($Q=16$, $S=8$) and \emcTwo\ attempted to recover the original image quality. The resulting images are compared in Figure \ref{fig:heart_comp}.

Due to the high contrast between the heart phantom and the surrounding water, gCNR is not an ideal metric for evaluating region separation. Accordingly, gCNR was replaced with edge-thickness measurements in this experiment. The wall and edge thickness of the heart was measured at three locations. Regions with values above -40 dB are considered part of the wall, whereas regions with values below -50 dB are considered background. Edge thickness is measured as the distance from the first peak over -40 dB to the first trough under -50 dB, while wall thickness is measured as the distance between the first and last peaks over -40 dB. These measurements are plotted in Figure \ref{fig:heart_measurements}. The motion-corrupted image exhibits significantly wider edges and walls than the static image, whereas the \emcTwo\ compensated image recovers sharper edges and accurate wall thicknesses. Speckle similarity was measured at the locations marked in green on Figure \ref{fig:heart_measurements}. It is collected in Table \ref{tb:heart_stats_transposed} along with the sharpness measurements and average thickness measurement errors. The recovered edge widths on the order of \textasciitilde 1 mm are comparable to the lateral resolution of \ac{forces} at those depths (presented in Appendix III-A).

To assess the effectiveness of the motion tracking algorithm, each READI image is compared with its neighbour rather than a single reference. Figure \ref{fig:motion_sample} shows the motion field between images 4 and 5 in the region near the servo connection, along with the correlation score for each vector. The colour of the motion field background indicates the total speed in mm/s. Under this, the measured average lateral velocity at the left edge is plotted from a set of 16 \ac{forces} images, split into 128 READI images. A clear sinusoidal pattern is visible as the heart beats, with an RMS average value of $44.8 \pm 1.4$ mm/s and a period of $0.502 \pm 0.019$ s ($119 \pm 4$ BPM), in good agreement with the estimated ground truth averages of 47.7 mm/s and 120 BPM. Supplementary figures are included in Appendix III-C that display all 8 READI low-resolution images and the 7 motion fields detected between them.

The correlation estimator successfully outputs a smooth motion field on the top and left portions of the heart. On the lower wall, however, the correlation score decreases, and the field becomes more random. The lower SNR due to attenuation and the increased READI artifacts at lower depths combine to produce noisier correlation surfaces, which strain the simple peak-detection algorithm.

An example cross-plane video of the beating heart phantom is included in the supplementary materials, comparing the original FORCES sequence to READI images without motion compensation. This is discussed further in section \ref{sec:volumetric_imaging}

\begin{table}
    \centering
    \begin{tabular}{|l||l|l|l|} 
         \hline
          Metric & \makecell{Static\\FORCES} & \makecell{Blurred\\FORCES} & \makecell{\emcTwo\\Compensated}\\
         \hline \hline
         Sharpness & 0.099 & 0.046 (-53.3\%)& 0.094 (-5.2\%)\\
         \hline
         \makecell[l]{Speckle\\Similarity} & N/A & 0.688 & 0.812\\
         \hline
         \makecell[l]{Average\\Wall Error} & N/A & 59.0\% & 11.1\%\\
         \hline
         \makecell[l]{Average\\Edge Error} & N/A & 316.4\% & 32.5\%\\
         \hline
    \end{tabular} 
    \vspace{0.5em}
    \caption{Beating Heart Phantom \emcTwo\ Performance }
    \label{tb:heart_stats_transposed}
\end{table}

\subsection{Flow Phantom}
As a final proof of concept, a 6 mm diameter blood vessel phantom (\textit{ATS 524}, CIRS Inc., Norfolk, VA, US) was imaged with an 8 MHz 128x128 element \ac{tobe} array at a \ac{prf} of 3.5 kHz and maximum depth of 25 mm. Blood-mimicking fluid was pumped through the phantom at a rate of 42 cm/s. \ac{readi} was performed with sixteen groups of eight transmits ($Q=8$, $S=16$) to improve blood speckle visibility. The results are displayed in Figure \ref{fig:flow_comp}. At these flow rates, scatterers can move by more than 12 mm during the FORCES sequence, preventing coherent compounding of the blood signal during beamforming. This results in a FORCES image in which the vessel appears nearly empty, with no speckle present other than the vessel walls. In contrast, the scatterer will move only 0.75 mm during an 8-transmit READI group, reducing negative interference and enabling stronger blood speckle.

The trade-off of using smaller READI groups of 8 transmits is noisier low-resolution images with more significant \ac{readi} artifacts. To remove these artifacts and isolate the flow signal, an ensemble of 16 \ac{forces} images was acquired, and each was beamformed with \ac{readi} to yield 256 low-resolution images in total. These images were clutter-filtered using basic \ac{svd} \cite{svd_clutter_review}. When applying SVD to the READI sequence in figure \ref{fig:flow_comp}, the largest singular value was kept to preserve the vessel walls for visualization purposes. 

After clutter filtering, the blood speckle is clearly visible as it flows through the vessel, and features of parabolic flow can be observed. The right side of Figure \ref{fig:flow_comp} tracks three speckle grains across four images. The motion of the grains is readily tracked between frames, and their speed differences due to parabolic flow are evident. The 42 cm/s flow meets or exceeds peak flow velocities in almost all veins, and is comfortably in the range of average larger arterial flow rates \cite{flow_rate_reference}.

In addition to directly filtering the 256 READI frames, they were combined into 241 FORCES images using a 16-frame moving window, after which the same SVD filter was applied. While flow dynamics are observable, individual speckle grains are smeared out by the compounding process and are not easily tracked. A video comparison of the original FORCES sequence to the clutter-filtered READI and moving window FORCES sequences is included in the supplementary materials. Both clutter-filtered sequences were superimposed on top of their unfiltered backgrounds for visualization purposes. This video has been slowed to make blood flow easier to track.

\begin{figure*}
    \centering
    \includegraphics[width=\linewidth]{figures/flow/blood_flow.pdf}
    \caption{6 mm diameter blood vessel phantom imaged with READI using 16 groups of 8 transmits. Top Left: Standard FORCES Image. Bottom Left: 8 Transmit READI image post-clutter filtering. Right: Four frames from the filtered READI sequence, with three speckle grains tracked.}
    \label{fig:flow_comp}
\end{figure*}
\begin{figure}
    \centering
    \includegraphics[width=\linewidth]{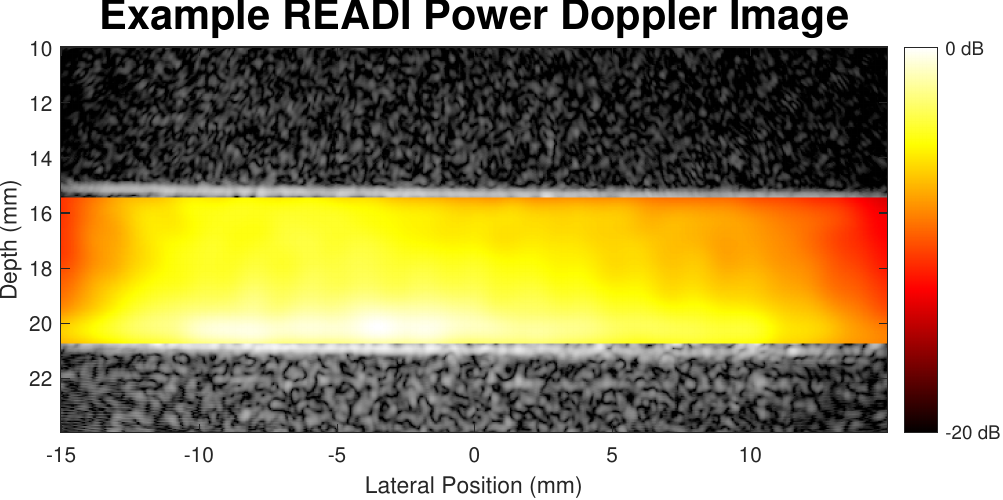}
    \caption{Example power Doppler image of the flow phantom, created by averaging the SVD clutter-filtered READI low-resolution images.}
    \label{fig:pd_example}
\end{figure}

\section{Discussion}
\subsection{Next Steps}

These results are extremely promising for the application of \ac{readi} to many ultrasound imaging tasks and provide clear next steps for further improvement. \\

\subsubsection{Motion Detection Algorithms}

The motion field shown in Figure \ref{fig:motion_sample} illustrates the limitations of the basic \ac{ncc} block-matching algorithm used in this work. While it successfully tracks motion along the top and left walls of the heart, its performance degrades on the lower wall, as lower SNR and more pronounced READI artifacts reduce image correlation. Once this correlation falls low enough, a large discontinuity in the field occurs, and the vectors begin to point in a spurious direction. Meanwhile, correlation scores on the left wall drop even lower, but the field remains smooth and points in the correct direction. This demonstrates that simple correlation with a threshold may not be the optimal method for \emcTwo\ motion detection. 

These issues may be mitigated by implementing more robust motion detection algorithms, such as biased block matching with priors \cite{cross_c_with_priors} or optical flow-based methods that require local motion continuity \cite{optical_flow_nerves}. A quantitative, no-reference cost function should be developed alongside each motion detection algorithm to automatically tune parameters without a priori knowledge or user input, starting with the gCNR and sharpness metrics used in this work. \\

\subsubsection{Real-Time Implementation}

Palamar \textit{et al.} \cite{forces_beamforming_preprint} present an open-source beamformer for \ac{rca} and \ac{tobe} arrays. They provide an in-depth analysis of the computational requirements for Hadamard decoding, envelope detection, and DAS beamforming. They demonstrate that consumer-grade GPUs can beamform \ac{forces} images with \textasciitilde 500k pixels at a rate of \textasciitilde 60 FPS. This is well above the current acquisition limit of \textasciitilde 24 FPS from the biasing hardware and is also limited by data transfer rates out of the Vantage system.

Applying \ac{readi} multiplies the FPS by the chosen number of groups without increasing the data transfer or acquisition rate. This allows us to bypass the identified bottlenecks, enabling real-time up reconstruction up to 60+ FPS. A \ac{readi} image requires the same amount of delay-and-sum operations as a \ac{forces} image, as each of the $Q$ signals is beamformed $S$ times.

\emcTwo\, with its current correlation-based algorithm, is too computationally expensive for real-time playback. The CUDA-based implementation of \emcTwo\ is roughly 20 times slower per image than beamforming, which would limit the maximum overall frame rate to below 3 FPS. As new motion detection algorithms are explored, one will be selected for low computational complexity and implemented in CUDA to enable real-time \emcTwo\ compensation. \\

\subsubsection{READI Artifact Filtering}
READI artifacts in the low-resolution images (Figure \ref{fig:cyst_low_res}) manifest as global intensity variations that worsen with smaller group sizes, thereby lowering the correlation between sequential images and reducing the effectiveness of motion compensation. Methods to filter out these artifacts should be explored to improve READI's performance with small group sizes. 

Testing with SVD filtering during the flow phantom experiment revealed that the majority of READI artifacts are mapped to singular values between the tissue and flow bands \cite{adaptive_svd}. This can potentially allow for selective filtering of these artifacts while maintaining both tissue and flow information. As a tradeoff, SVD is computationally expensive and requires a large ensemble of images to be effective. Another option is applying a method similar to flat-field correction in optical imaging, where the background brightness pattern is modelled and then divided out of the image \cite{flat_field_correction}. 

\subsection{Future Work}

\subsubsection{Volumetric Imaging with READI} \label{sec:volumetric_imaging}

\begin{figure}
    \centering
    \includegraphics[width=\linewidth]{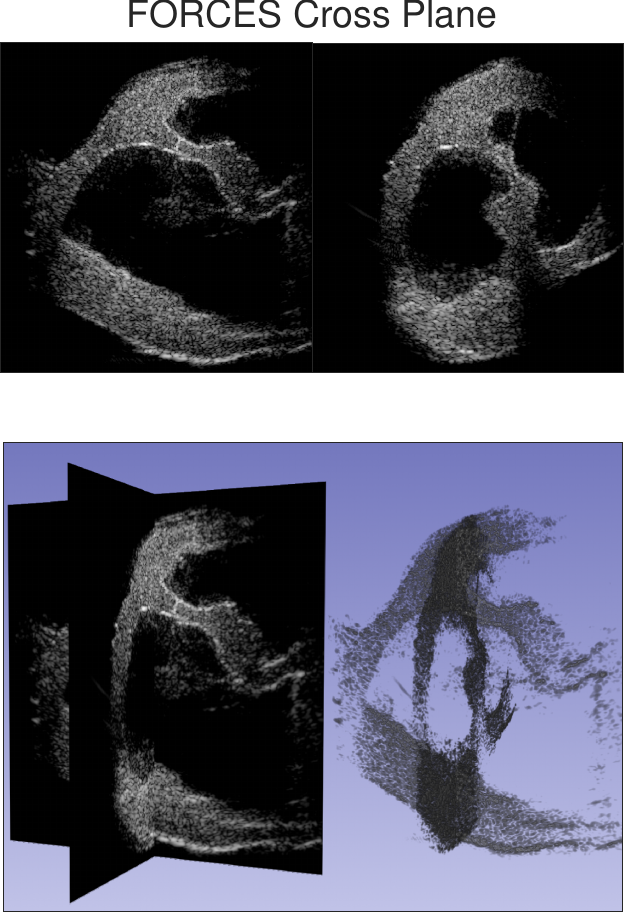}
    \caption{Example cross-plane volume of the heart phantom, demonstrating the 3D capabilities of \ac{forces}}
    \label{fig:cross_plane_example}
\end{figure}

\ac{readi} and \emcTwo\ can be directly applied to the walking and cross-plane FORCES sequences discussed in the background, where READI is applied independently to each FORCES plane. Figure \ref{fig:cross_plane_example} provides an example cross-plane image of the heart phantom, and a fly-around video of crossed planes is provided in the supplementary materials. Additionally, a video of both planes beating simultaneously is also provided. In this video, the low-frame-rate FORCES sequence is compared with the high-frame-rate READI 32 sequence without motion compensation. The motion in the READI sequence is noticeably smoother than the FORCES sequence, and the loss in quality due to lower transmit counts is offset by the decreased motion blur.

The example cross-plane sequence was taken with a PRF of 1 kHz. As it interleaves two FORCES planes, each plane has an effective PRF of 0.5 kHz. With the current PRF limit of \textasciitilde 3 kHz, six FORCES sequences on different planes could be interleaved in the same manner. Each plane would have an effective PRF of 0.5 kHz, so its motion corruption would be comparable to the \emcTwo\ heart example (Figure \ref{fig:heart_comp}). The width of these images (-45 to +45 mm) extends far beyond the shadow of the aperture (-16 to +16 mm). This is unlike most previous RCAs \cite{forces_comp_preprint}. \emcTwo\ combined with this large field-of-view enables 3D tissue/organ characterization in the presence of motion, such as chamber volume estimation. 

Beyond FORCES, the recursion and partial decoding at the core of READI applies to all Hadamard aperture-encoded sequences. \ac{hercules} is a recently presented 3D Hadamard encoded sequence on \ac{tobe} arrays \cite{hercules_preprint} that uses the same amount of transmission events as 2D FORCES. Rather than generating a 2D image with global transmit and receive focusing, it produces a 3D volume with global receive focusing in all directions. This technique suffers from the same motion sensitivity sources as \ac{forces}, so it can benefit from \ac{readi} and \ac{emc2} to reduce motion artifacts. \\

\subsubsection{Flow Estimation}
The success of the flow phantom experiment suggests that flow imaging techniques could be applied to \ac{readi} images. While some vessels experience flow rates exceeding 42 cm/s, the simulations in Appendix II-B suggest that, at a PRF of 3.5 kHz, coherent lateral motion is measurable at speeds up to 188 cm/s. Currently, measurable flow velocities may be insufficient to measure pathological conditions such as carotid stenoses. New biasing electronics with higher PRF capabilities would allow for even higher flow rates to be measured. The clutter filtered READI sequence is already close to a power Doppler image, and the clear speckle flow visible in it suggests that more robust colour and vector flow techniques should be applicable \cite{vector_flow_review} \cite{transverse_oscillations}. As an example, a basic power Doppler image was created by time averaging the 256 SVD-clutter filtered READI images (Figure \ref{fig:pd_example}).

Applying READI to flow imaging can produce both a high-quality B-mode image and a detailed flow map. Both datasets would be perfectly co-registered in time without the need to interleave separate B-mode and Doppler sequences. The improved edge sharpness at tissue boundaries, enabled by motion compensation, should also improve the accuracy of flow characterization by providing more precise vessel/chamber boundaries. \\

\subsubsection{Elastography}
Another imaging modality that involves motion is shear-wave elastography. This technique induces a shear wave in tissue and tracks its propagation via high-frame rate B-mode imaging \cite{shear_wave_review}. Shear waves travel and attenuate rapidly in soft tissue, requiring an imaging method with a high frame rate while still producing images detailed enough to track subtle tissue deformations as the shear wave propagates. The comparison of \ac{readi} images with \ac{uforces} images of the same transmit count suggests that \ac{readi} may be ideal for this application, as it maintains good structural resolution over a large field-of-view at low transmit counts. Shear-wave elastography has previously been implemented in 3D with standard row-column arrays \cite{shear_wave_rca}, so it is a good next step for \ac{readi} on \ac{tobe} arrays.

\section{Conclusion}
This work presents \acreset{readi}\ac{readi} with \acreset{emc2}\ac{emc2}, a new method for beamforming Hadamard aperture-encoded sequences that reduces motion artifacts while maintaining high spatial resolution. The intermediate \ac{readi} low-resolution images are shown to be a marked improvement over sparse transmit aperture \ac{uforces} images at the same transmission count. \ac{readi} with \ac{emc2} successfully removed blur due to probe motion, reproducing an image of comparable quality to standard \ac{forces} with a static probe. With a beating heart phantom, \ac{readi} \ac{emc2} removed speckle blur and sharpened the edges of moving tissue regions, showing its potential for tissue motion cancelling and region segmentation. \ac{readi} also successfully recovered blood speckle in a vessel phantom at a flow speed of 42 cm/s, demonstrating its potential for high-resolution flow imaging in large arteries and veins.
Multiple avenues for further development of \ac{readi} have been identified, including 3D volumetric imaging, vector flow characterization, and shear-wave elastography. These applications will be explored in future work.

\section{Conflict of Interest Statement}
The authors R.J.Z. and M.R.S. are directors and shareholders of CliniSonix Inc., which provided partial support for this work. RJZ is also a founder and director of OptoBiomeDx Inc., and a founder and shareholder of IllumiSonics Inc., however neither organization provided support for this work. 

\newpage
\printacronyms[template=tabular,display=used]

\bibliographystyle{IEEEtran}
\bibliography{bib/references}

\newpage

\renewcommand{\thefigure}{A.\arabic{figure}}
\setcounter{figure}{0}

\renewcommand{\thetable}{A.\Roman{table}}
\setcounter{table}{0}


\appendices
\renewcommand{\theequation}{\thesection.\arabic{equation}}
\makeatletter
\@addtoreset{equation}{section}
\makeatother

\section{READI Derivation} \label{derivation_appendix}
This section provides a detailed derivation of the \acs{readi} reconstruction operator (eq. 12) from the base \acs{forces} reconstruction operator (eq. 6):
\begin{gather*}
    \text{Starting point:} \\
    \mathbb{F}\{ \sigmatm{G}(t),r\} = \mathbb{D}\{\bm{H}_N^{-1} \sigmatm{G}(t),r\} \\ \\
    \text{Goal:} \\
    \mathbb{R}\{  \sigmatm{g'}_s(t),r\} = \sum_{l}^{S} \hat{h}^{(S)}_{l,s} \mathbb{D'}_{l} \{ \bm{H}_Q^{-1} \sigmatm{g'}_{s}(t), r \} \\
    \mathbb{F}\{ \sigmatm{G'}(t),r\} = \sum_{s}^S \mathbb{R}\{  \sigmatm{g'}_s(t),r\}
\end{gather*}

In this work, $\bm{H}_N$ denotes a Hadamard matrix of rank $N$, and the following notation is used to reference individual elements of the matrix:
\begin{equation*}
    \bm{H}_N = [h^{(N)}_{ij}],\ \bm{H}_N^{-1}  = [\hat{h}^{(N)}_{ij}]
\end{equation*}

To review, the \acs{forces} dataset of $N$ encoded transmissions is split into $S$ groups of $Q$ transmissions. The transmit element $i$ and transmit event $e$ are reindexed to reflect this grouping:
\begin{equation}
\begin{aligned}
	i &= v + Q(l-1) \\
	e &= q + Q(s-1)
\end{aligned}
\end{equation}
Where $l$ and $s$ are the transmit element and event group numbers, while $v$ and $q$ are the element's and event's indices within their group. In this analysis, $(vl)$ is written in place of $(v+Q(l-1))$ and $(qs)$ is written in place of $(q+Q(s-1))$ for space.

To align with this grouping, the dataset $\sigmatm{G}(t)$ is reordered:
\begin{equation*}
\begin{array}{c}
\sigmatm{G}(t) \in \mathbb{R}^{N \times 1} \to \sigmatm{G'}(t) \in \mathbb{R}^{Q \times S} \\
\sigmatm{g'}_{s}(t) := \mathrm{col}_{s} \sigmatm{G'}(t) \  (s = 1, \ldots, S) \\
g'_{q,s}(t) = \sum_{l=1}^{S}\sum_{v=1}^{Q} h^{(N)}_{(qs),(vl)} s_{(vl)}(t)
\end{array}
\end{equation*}
Now \txgroup represents the $s$-th group of $Q$ transmit events.

Thanks to its recursive Kronecker product definition, the Hadamard matrix $\bm{H}_{N}$ can be linearly separated into two smaller Hadamard matrices $\bm{H}_{S}$ and $\bm{H}_{Q}$ provided $N$, $S$, and $Q$ are all powers of 2:
\vspace{0.5em}
\begin{equation}
    \begin{array}{c}
        \bm{H}_{N} = \bm{H}_S \otimes \bm{H}_Q \\[1.5ex]
        h^{(N)}_{ie} = h^{(N)}_{(vl),(qs)} = h^{(S)}_{s,l} h^{(Q)}_{q,v} \\[1.5ex]
        \bm{H}_N = [h^{(N)}_{ij}]
    \end{array}
\end{equation}

\vspace{0.5em}
Similarly, for its inverse:
\vspace{0.5em}
\begin{equation}
    \begin{array}{c}
        \bm{H}^{-1}_{N} = \bm{H}^{-1}_{S} \otimes \bm{H}^{-1}_{Q} \\[1.5ex]
        \hat{h}^{(N)}_{ei} = \hat{h}^{(N)}_{(qs),(vl)} = \hat{h}^{(S)}_{l,s} \hat{h}^{(Q)}_{v,q}
    \end{array}
\end{equation}
\vspace{0.5em}

Next, the reindexing and Kronecker definitions are applied to the \acs{forces} reconstruction equation; the matrix multiplication decode step is kept within curly braces:
\begin{align*} 
    \mathbb{F}\{ \sigmatm{G}(t),r\} &= \mathbb{D}\{\bm{H}_N^{-1} \sigmatm{G}(t),r\}\\
    \text{Expand} &\Rightarrow \sum_{i}^{N} \{\sum_{e}^{N} \hat{h}^{(N)}_{ie} g_{e}\}(\tau_{i}(r)) \\
    \text{Reindex} &\Rightarrow \sum_{l}^{S}\sum_{v}^{Q}\{\sum_{s}^{S}\sum_{q}^{Q} \hat{h}^{(N)}_{(vl),(qs)} g'_{q,s}\}(\tau_{(vl)}(r)) \\
    \text{Split } \bm{H}^{-1}_N  &\Rightarrow \sum_{l}^{S}\sum_{v}^{Q}\{\sum_{s}^{S}\sum_{q}^{Q} \hat{h}^{(S)}_{l,s} \hat{h}^{(Q)}_{v,q} g'_{q,s}\}(\tau_{(vl)}(r))\\
    \text{Reorder} &\Rightarrow \sum_{s}^{S}[\sum_{l}^{S}\hat{h}^{(S)}_{l,s}\sum_{v}^{Q}\{\sum_{q}^{Q}\hat{h}^{(Q)}_{v,q} g'_{q,s}\}(\tau_{(vl)}(r))]\\
    \mathbb{F}\{ \sigmatm{G'}(t),r\} &= \sum_{s}^{S}[\sum_{l}^{S}\hat{h}^{(S)}_{l,s}\sum_{v}^{Q}\{\bm{H}_Q^{-1} \sigmatm{g'}_{s}(t)\}(\tau_{(vl)}(r))] \numberthis \label{eq:a-expanded-forces}
\end{align*}

The matrix multiplication in curly braces is now a partial decoding of a single transmit group, and is defined as a \textbf{partially decoded dataset} \partDec:
\begin{gather} 
    \sigmatm{D}(t) \in \mathbb{R}^{Q \times S} \notag\\
    \sigmatm{d}_{s}(t) = \bm{H}_Q^{-1} \sigmatm{g'}_{s}(t) \notag\\
    d_{v,s}(t) = \sum_{q=1}^{Q} \hat{h}^{(Q)}_{v,q} g'_{q,s}(t) \label{eq:partially-decoded-def}
\end{gather}

As discussed in section II-B.2, the \acs{das} operator associates each signal with a unique transmit element for beamforming, but, in equation \ref{eq:a-expanded-forces}, \acs{das} must be applied to $Q$ partially decoded signals from an array with $N$ transmit elements. Like the events, the $N$ transmit elements are also divided into $S$ groups of $Q$. Thus, a modified \acs{das} operator is defined that associates an event group of $Q$ signals with the $l$'th group of elements:
\begin{equation}
    \mathbb{D'}_{l}\{\sigmatm{d}_s(t),r\} = \sum_{v=1}^{Q} d_{v,s}(\tau_{v + Q(l-1)}(r))
\label{eq:grouped-das}
\end{equation}
Note that while $s$ denotes the event group this data is taken from, $l$ is used to determine the transmit location for the delays.

Applying this new \acs{das} operator to equation \ref{eq:a-expanded-forces} produces the \acs{readi} reconstruction operator (eq. 12):
\begin{gather*}
\mathbb{F}\{  \sigmatm{G'}(t),r\} =  \sum_{s}^S [\sum_{l}^{S} \hat{h}^{(S)}_{l,s}\mathbb{D'}_{l} \{ \bm{H}_Q^{-1} \sigmatm{g'}_{s}(t), r \}] \\
\mathbb{R}\{  \sigmatm{g'}_s(t),r\} = \sum_{l}^{S} \hat{h}^{(S)}_{l,s} \mathbb{D'}_{l} \{ \bm{H}_Q^{-1} \sigmatm{g'}_{s}(t), r \} \\
\mathbb{F}\{ \sigmatm{G'}(t),r\} = \sum_{s}^S \mathbb{R}\{  \sigmatm{g'}_s(t),r\}
\end{gather*}
To form a single \acs{readi} low-resolution image from \txgroup, the modified \acs{das} operator is called $S$ times, once for each element group $l$. Each DAS iteration produces a sub-image, and the \acs{readi} image is formed as the sum of these sub-images weighted by the $s$'th row of $\bm{H}_S^{-1}$. 

Appendix \ref{operator_appendix} analyzes the partially decoded dataset $\sigmatm{d}_{s}(t)$ to understand how it differs from a standard decoded \acs{forces} dataset and explains the presence of $\hat{h}^{(S)}_{l,s}$ in the reconstruction operator.

\section{READI Reconstruction Operator Analysis} \label{operator_appendix}
To understand how the \acs{readi} reconstruction operator recovers the original \acs{forces} image we must analyze what happens when we attempt to beamform a group from the partially decoded dataset \partDecGroup\ using standard delay-and-sum. First, \partDecGroup\ must be defined in relation to the multistatic dataset $\sigmatm{S}(t)$. Substituting the definition of $g'_{q,s}(t)$ (eq. 10) into equation \ref{eq:partially-decoded-def}:
\begin{align*}
d_{v,s}(t) &= \sum_{q=1}^{Q} \hat{h}^{(Q)}_{v,q} g'_{q,s}(t) \\
\text{Expand}\ g'_{q,s}(t) &\Rightarrow \sum_{q}^{Q} \hat{h}^{(Q)}_{v,q} \sum_{l=1}^{S}\sum_{v'=1}^{Q} h^{(N)}_{(qs),(v'l)} s_{(v'l)}(t) \\
\text{Split}\ \bm{H}_N &\Rightarrow \sum_{q}^{Q} \hat{h}^{(Q)}_{v,q} \sum_{l}^{S}\sum_{v'}^{Q} h^{(S)}_{s,l} h^{(Q)}_{q,v'} s_{(v'l)}(t) \\
\text{Reorder} &\Rightarrow \sum_{l}^{S}\sum_{v'}^{Q} h^{(S)}_{s,l} s_{(v'l)}(t) \sum_{q}^{Q} \hat{h}^{(Q)}_{v,q} h^{(Q)}_{q,v'}\\ 
\bm{H}_Q \bm{H}_Q^{-1} = \bm{I}_Q &\Rightarrow  \sum_{l}^{S}\sum_{v'}^{Q} h^{(S)}_{s,l}  s_{(v'l)}(t) \delta(v,v') \\
d_{v,s}(t) &= \sum_{l}^{S} h^{(S)}_{s,l}  s_{v+Q(l-1)}(t) \numberthis \label{eq:partially-decoded}
\end{align*}

Thus, the $v$'th signal in the partially decoded dataset $\sigmatm{d}_{s}(t)$ is a linear combination of the contributions from the $v$'th transmit element of every group. This is very visible in Figure 2C, where with four groups of two events ($N=8\ S=4\ Q=2$) the partially decoded apertures contain $S=4$ signals apiece. In general, the number of terms mixed in each partially decoded signal $d_{v,s}(t)$ is equal to the number of groups $S$, and their coefficients are given by the $s$-th row of $\bm{H}_S$.

We will apply this definition to the case of two groups of two transmit events each ($N=4\ S=2\ Q=2$):
\begin{gather*}
\bm{H}_S = \begin{bmatrix}
    1 & 1 \\
    1 & -1
\end{bmatrix} \\
\sigmatm{d}_{1}(t) = \begin{bmatrix}
    s_1(t) + s_3(t)\\
    s_2(t) + s_4(t)
\end{bmatrix}\ 
\sigmatm{d}_{2}(t) = \begin{bmatrix}
    s_1(t) - s_3(t)\\
    s_2(t) - s_4(t)
\end{bmatrix}
\end{gather*}

Both partially decoded groups contain all four multistatic signals, mixed according to the group's row in the $\bm{H}_S$ ($\bm{H}_2$) matrix. Attempting to directly beamform these groups using the \acs{das} operator (equation 5) would produce images with significant artifacts. Per equation 6, the correct reconstruction of $\sigmatm{G}(t)$ for an $N=4$ \acs{forces} sequence is:
\begin{equation*} 
    \mathbb{F}\{  \sigmatm{G}(t),r\} = s_1(\tau_1) + s_2(\tau_2) + s_3(\tau_3) + s_4(\tau_4)
\end{equation*}
Where the contribution from each transmitting element $s_i(t)$ is delayed according to that element's position $\tau_i$, bringing the point into focus. However, attempting standard \acs{das} beamforming on $\sigmatm{d}_{1}(t)$ yields:
\begin{equation*}
    \mathbb{D}\{\sigmatm{d}_{1}(t),r\} = s_1(\tau_1) + s_2(\tau_2) + s_3(\tau_1) + s_4(\tau_2)
\end{equation*}
The elements of group one ($s_1$ and $s_2$) are beamformed with correct delays, whereas the second group ($s_3$ and $s_4$) is incorrectly delayed and out of focus. To obtain all four in-focus signals $\sigmatm{d}_{1}(t)$ must be beamformed twice with the modified \acs{das} operator (\ref{eq:grouped-das}), once for each element group:
\begin{gather*}
\mathbb{D'}_{2}\{\sigmatm{d}_{1}(t),r\} = s_1(\tau_3) + s_2(\tau_4) + s_3(\tau_3) + s_4(\tau_4)
\end{gather*}
\begin{multline} \label{eq:readi_group_1}
       \mathbb{R}\{\sigmatm{g'}_{1}(t),r\} =\mathbb{D'}_{1}\{\sigmatm{d}_{1}(t),r\} + \mathbb{D'}_{2}\{\sigmatm{d}_{1}(t),r\} = \\
        s_1(\tau_1) + s_2(\tau_2) + s_3(\tau_3) + s_4(\tau_4) \\
        + s_1(\tau_3) + s_2(\tau_4) + s_3(\tau_1) + s_4(\tau_2)
\end{multline}
This produces an image that contains all in-focus data present in the full \acs{forces} image, but with significant artifacts from extra cross terms. We refer to this image as a \textbf{READI low-resolution image}, while the individual outputs of the $\mathbb{D'}_{l}\{\sigmatm{d}_{s}(t),r\}$ operators are referred to as \textbf{READI sub-images}.

A second issue arises when we attempt to beamform $\sigmatm{d}_{2}(t)$. As its components are weighted by the second row of $\bm{H}_2$ ($[1,-1]$) all terms from the second group of elements (3 and 4) are negative. To counteract this, we subtract the second sub-image from the first when forming the \acs{readi} low-resolution image:
\begin{multline} \label{eq:readi_group_2}
       \mathbb{R}\{\sigmatm{g'}_{2}(t),r\} = \mathbb{D'}_{1}\{\sigmatm{d}_{2}(t),r\} - \mathbb{D'}_{2}\{\sigmatm{d}_{2}(t),r\} = \\
        s_1(\tau_1) + s_2(\tau_2) + s_3(\tau_3) + s_4(\tau_4) \\
        - s_1(\tau_3) - s_2(\tau_4) - s_3(\tau_1) - s_4(\tau_2)
\end{multline}
In general, when forming the \acs{readi} low-resolution image for the $s$-th group, the sub-images are weighted according to the $s$-th column of $\bm{H}_S^{-1}$, which ensures that all in-focus terms have the same sign. Adding this weighting ($\hat{h}^{(S)}_{l,s}$) completes the definition of the \acs{readi} reconstruction operator:
\begin{equation*}
\mathbb{R}\{  \sigmatm{g'}_s(t),r\} = \sum_{l}^{S} \hat{h}^{(S)}_{l,s} \mathbb{D'}_{l} \{ \bm{H}_Q^{-1} \sigmatm{g'}_{s}(t), r \}
\end{equation*}

Both \acs{readi} low-resolution images now contain all in-focus terms, along with a set of out-of-focus cross-terms that degrade image quality. When these two low-resolution images are summed, these cross-terms cancel out, leaving only the original \acs{forces} image:
\begin{align*}
    \mathbb{F}\{  \sigmatm{G'}(t),r\} &= \mathbb{R}\{\sigmatm{g'}_{1}(t),r\} + \mathbb{R}\{\sigmatm{g'}_{2}(t),r\} \notag \\
    &= s_1(\tau_1) + s_2(\tau_2) + s_3(\tau_3) + s_4(\tau_4)
\end{align*}

\newpage

\section{Additional Results} \label{results_appendix}

\subsection{Point Spread Function Simulations} \label{sec:ap_psf_simulation}

\begin{figure}
    \centering
    \includegraphics[width=\linewidth]{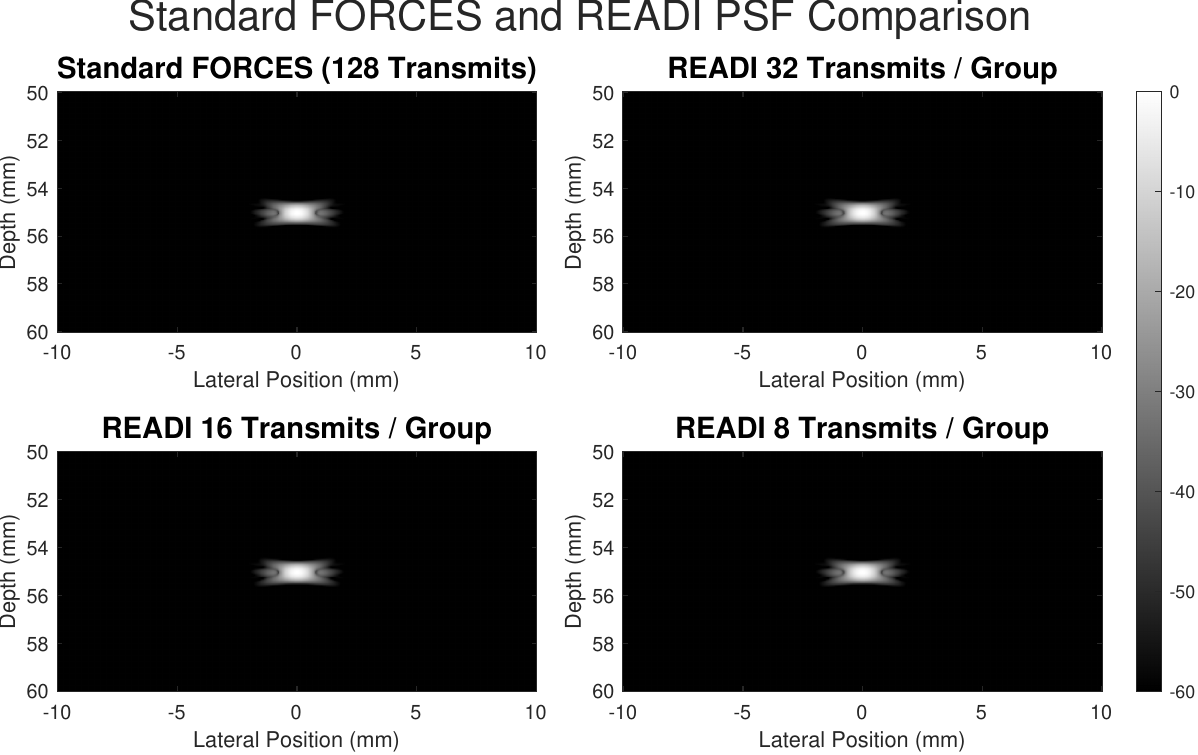}
    \caption{Point Spread Function simulation in Field II, comparing standard \acs{forces} beamforming with \acs{readi} beamforming using groups of 32, 16, and 8 transmits. All images are identical within numerical precision.}
    \label{fig:psf_readi_forces_comp}
\end{figure}

\begin{figure}
    \centering
    \includegraphics[width=\linewidth]{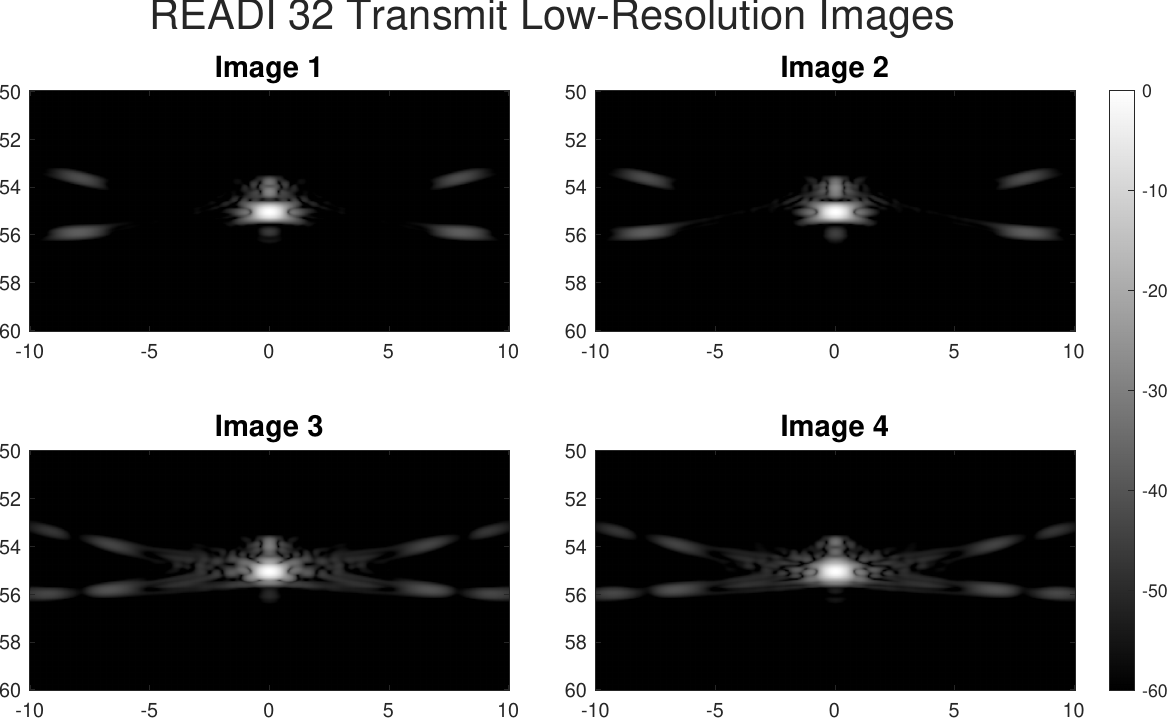}
    \caption{READI low-resolution point spread functions formed from 32 transmits apiece. Coherently summing these four images produces the top right image in Figure \ref{fig:psf_readi_forces_comp}.}
    \label{fig:psf_low_res}
\end{figure}

Coherence-factor weighting was disabled in these simulations to preserve linearity.

To validate the \acs{readi} beamforming process, a 128-transmit \acs{forces} \acs{psf} was simulated in Field II. The recorded data was beamformed four times, once using the standard \acs{forces} beamformer and three times using \acs{readi}, with group sizes ($Q$) of 32, 16, and 8 transmits. Figure \ref{fig:psf_readi_forces_comp} displays the beamformed images. In all cases, the \acs{readi} beamformer successfully reproduces the static \acs{forces} image, with average per-pixel errors on the order of $10^{-3}$ dB at a dynamic range of 100 dB. This error is attributable to 32-bit floating-point precision limitations and is negligible compared to the noise floor of real acquisitions.

Figure \ref{fig:psf_low_res} displays the four low-resolution \acs{readi} PSFs formed with $Q=32$. Each \acs{psf} is centred at the same position but has its own set of unique artifacts. These \acs{readi} artifacts are generated by the cross-terms in the partially decoded dataset (described in Appendix \ref{operator_appendix} equations \ref{eq:readi_group_1} and \ref{eq:readi_group_2}). Each group has a distinct set of artifacts due to the different signs of its cross-terms; when all images are summed, these artifacts cancel out. 
\begin{figure}
    \centering
    \includegraphics[width=\linewidth]{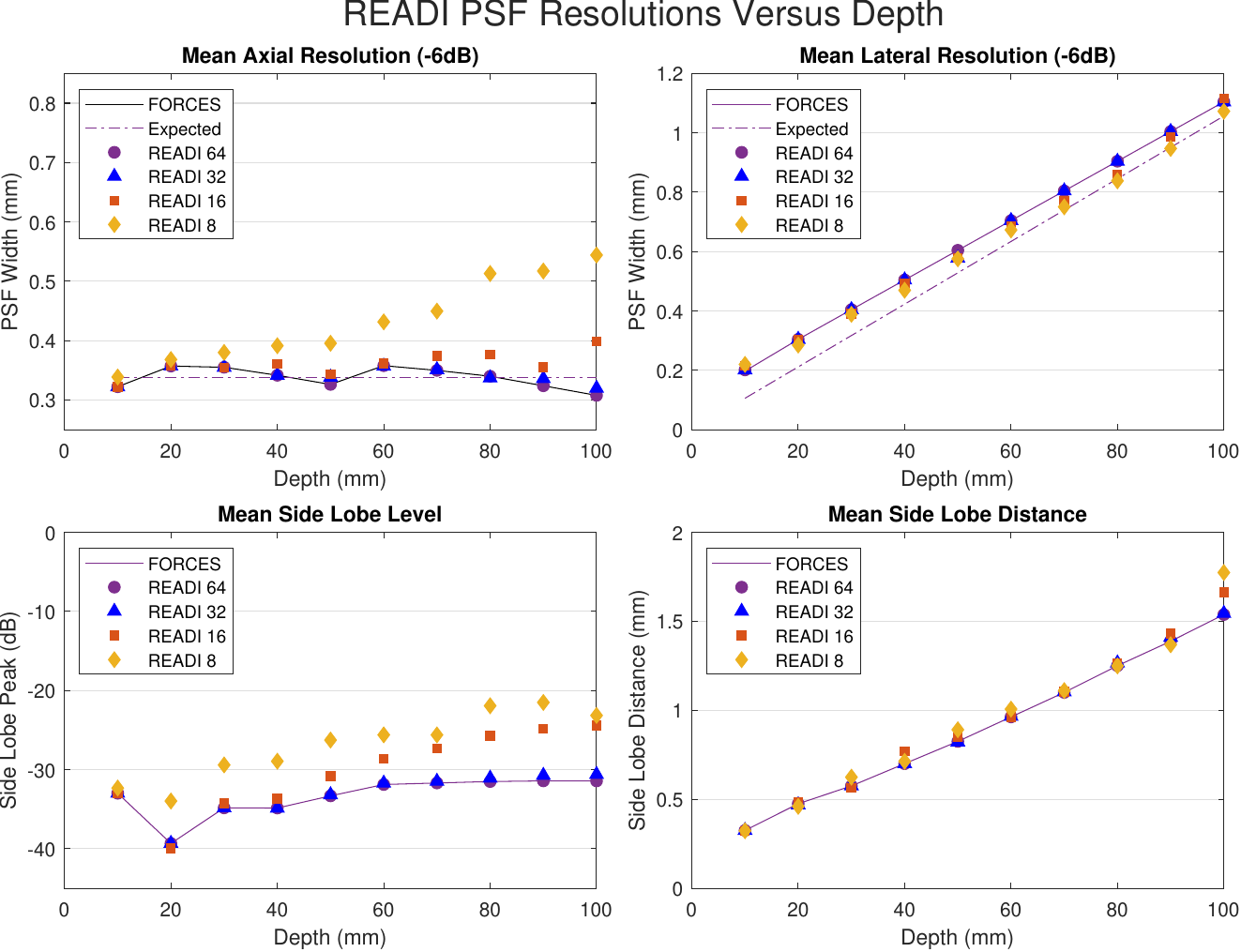}
    \caption{Field II PSF resolution measurements for FORCES and READI with different group sizes. All \acs{readi} results are mean values from the low-resolution images.}
    \label{fig:reso_sim}
\end{figure}

\begin{figure*}
    \centering
    \includegraphics[width=\linewidth]{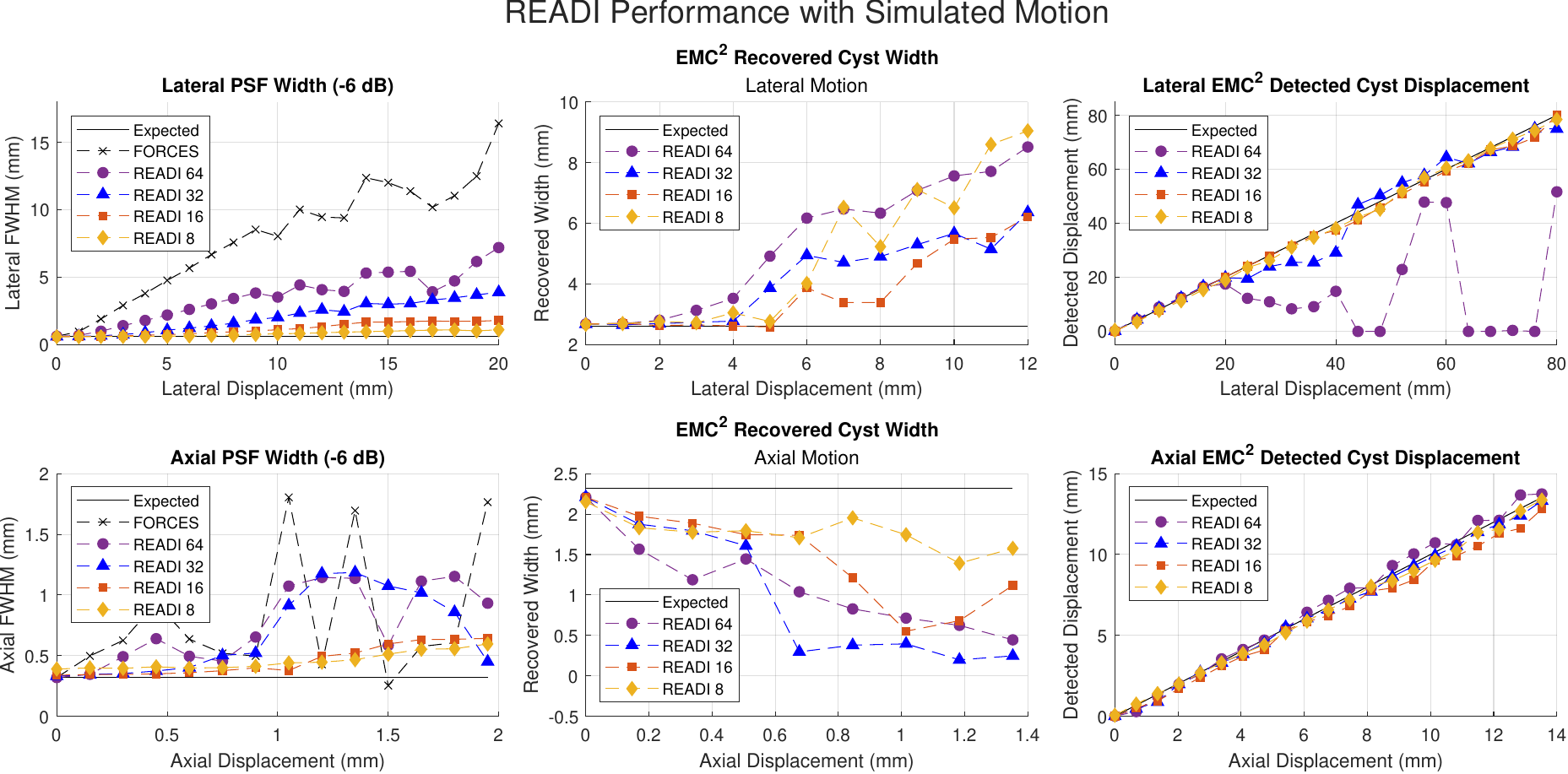}
    \caption{Field II READI performance with axial and lateral motion. Displacement values state the total amount of motion over the 128 transmit FORCES sequence.}
    \label{fig:motion_sim}
\end{figure*}

To analyze the resolution characteristics, the stationary FORCES \acs{psf} was simulated at various depths and beamformed with multiple READI group sizes (64, 32, 16, and 8 transmits). The axial and lateral resolution was measured at each depth, along with sidelobe position and height. The results are averaged across all READI images for each group size. Results are presented in Figure \ref{fig:reso_sim}. All group counts have similar lateral resolution, closely following the FORCES image and the expected resolution of $\lambda F_\#$ for a wavelength of 330 $\mu m$ and a 32 mm wide array. Axially, all cases stay close to the expected value of $\lambda$ for a fully focused 2-cycle pulse; however, the reduced transmit focusing begins to have noticeable effects on READI 8 at lower depths. Since each READI 8 image only contains 8 transmissions, this is an understandable loss of quality. We also observe that the number of READI groups has no effect on the nearest sidelobe position (defined as the first lateral peak in the PSF off-center); however, as the READI group count increases, sidelobe heights can also increase. Table \ref{tb:psf_sim_params} lists all relevant simulation parameters for this work and the simulation results presented in the main text. All simulation code is available upon request.\\

\begin{table}
    \centering
    \begin{tabular}{||c|c||} 
        \hline
        Array Type & 128x128 TOBE\\
        \hline
        Transmit Frequency & 4.3 MHz \\
        \hline
        Element Pitch & 0.25 mm \\
        \hline
        Transmit Waveform & 2-cycle sinusoid \\
        \hline
        Sampling Frequency & 50 MHz \\
        \hline
        Speed of Sound & 1540 m/s \\
        \hline
        Elevational Focal Depth & 55 mm \\
        \hline
        Transmit Apodization Pattern & Hanning \\
        \hline
    \end{tabular} 
    \vspace{0.5em}
    \caption{Field II PSF Simulation Array Parameters}
    \label{tb:psf_sim_params}
\end{table}
\vspace{0.5em}

\subsection{Motion Simulation} \label{ap:motion_sim}

Two simulations were performed to analyze axial and lateral performance, using the same parameters as in the resolution analysis above. First, a PSF is simulated moving at constant axial and lateral rates, and its resolution in the relevant direction is measured. Then, a 2 mm-diameter spherical hyperechoic cyst is moved against a uniform background of speckle. Its motion was estimated with NCC, and \emcTwo\ attempted to reconstruct the cyst. 30 dB of white noise was added to the simulated RF data to more closely approximate real-world conditions. All test cases are expressed as displacements (the distance the point moves over the 128 FORCES transmissions). These results are collected in Figure \ref{fig:motion_sim}. The simulation settings for the PSFs and the array are the same as provided in Table \ref{tb:psf_sim_params}. Table \ref{tb:cyst_sim_params} describes the scatterer parameters for the cyst simulation. 

\begin{table}
    \centering
    \begin{tabular}{||c|c||} 
        \hline
        Number of Scatterers & 280902 \\
        \hline
        Lateral Scatterer Range & -45 to 45 mm \\
        \hline
        Elevational Scatterer Range & -1 to 1 mm \\
        \hline
        Axial Scatterer Range &  40 to 60 mm \\
        \hline
        Cyst Diameter & 2 mm \\
        \hline
        Cyst Shape & Spherical \\
        \hline
        Cyst Scatterer Strength & 15x Background Strength \\
        \hline
    \end{tabular} 
    \vspace{0.5em}
    \caption{Field II Cyst Simulation Scatterer Parameters}
    \label{tb:cyst_sim_params}
\end{table}
\vspace{0.5em}

Starting with the PSF measurements on the left of Figure \ref{fig:motion_sim}, we observe that the measured PSF width scales proportionally with the amount of lateral motion. As the sequence is divided into READI groups, the slope of this curve is also divided by the number of groups. Axially, this trend is less stable. In the FORCES case, we see an abrupt spike in PSF width as soon as motion begins. With READI 2, this spike is shifted further out, and with smaller group sizes, it smooths into a gradual increase. In both cases, the measured resolution remains significantly more stable at higher READI group counts. 

For the cyst motion simulation (center and right), we observe that \emcTwo\ fails to recover the correctly sized cyst at 10 mm of lateral displacement and 1 mm of axial displacement (over 128 transmissions), yet the measured motion fields remain accurate far past these cutoffs. READI accurately measures cyst motion with up to 80 mm of lateral displacement and up to 14 mm of axial displacement. At a PRF of 1 kHz, these correspond with speeds of 62.5 cm/s and 11.0 cm/s, respectively; at 3.5 kHz, they scale to 187.5 cm/s and 48.5 cm/s. These extremely high speeds indicate that, even after \emcTwo\ fails to reconstruct the static image, READI is still capable of delivering accurate motion estimates. All simulation code is available upon request.

\subsection{Heart Phantom Results} \label{ap:heart}

\begin{figure*}
    \centering
    \includegraphics[width=\linewidth]{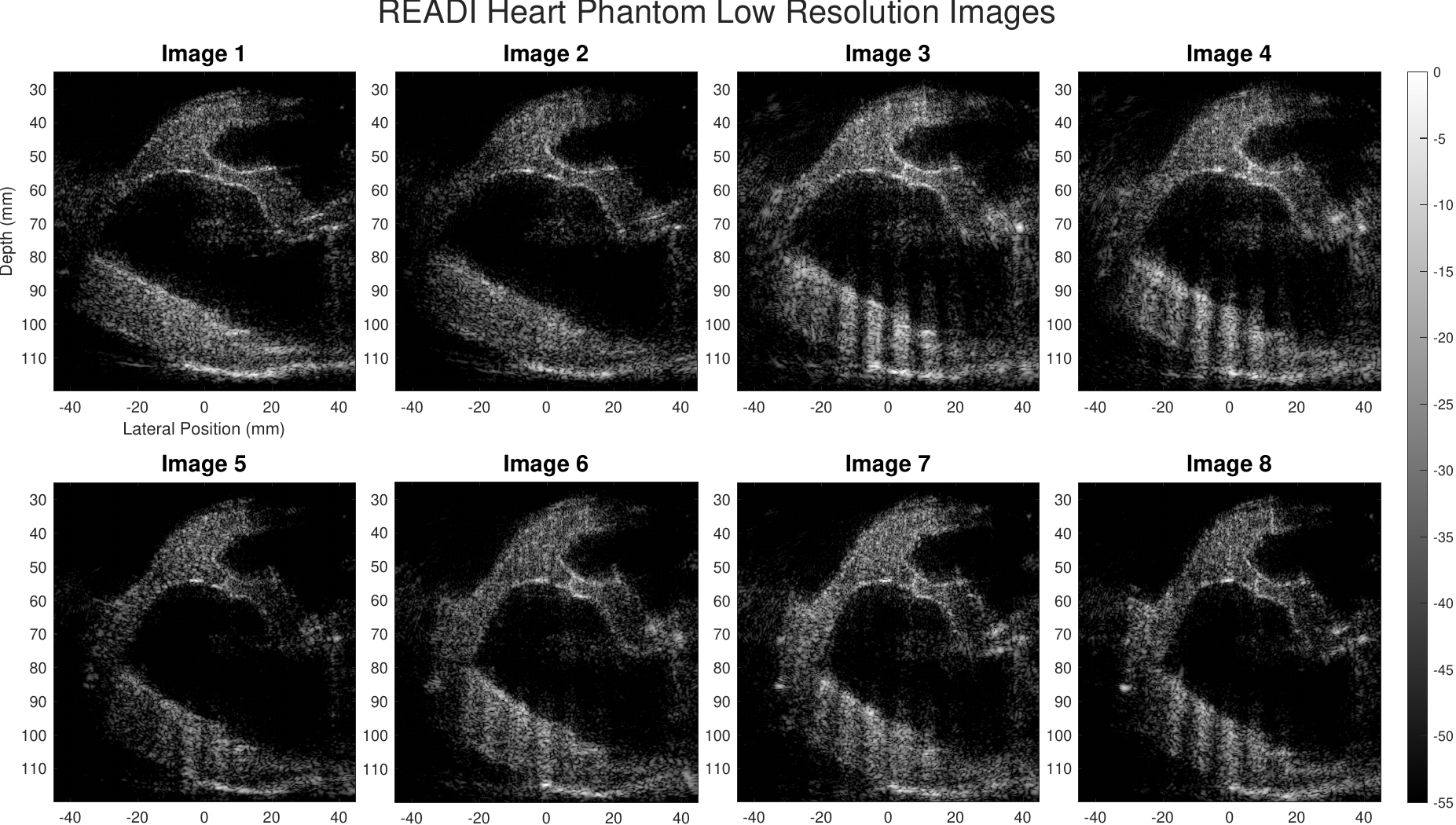}
    \caption{The 8 READI low-resolution images from the heart phantom experiment. These images in the input into \emcTwo\ and produce the results in Figures 8 and 9}
    \label{fig:all_heart_lris}
\end{figure*}

\begin{figure*}
    \centering
    \includegraphics[width=\linewidth]{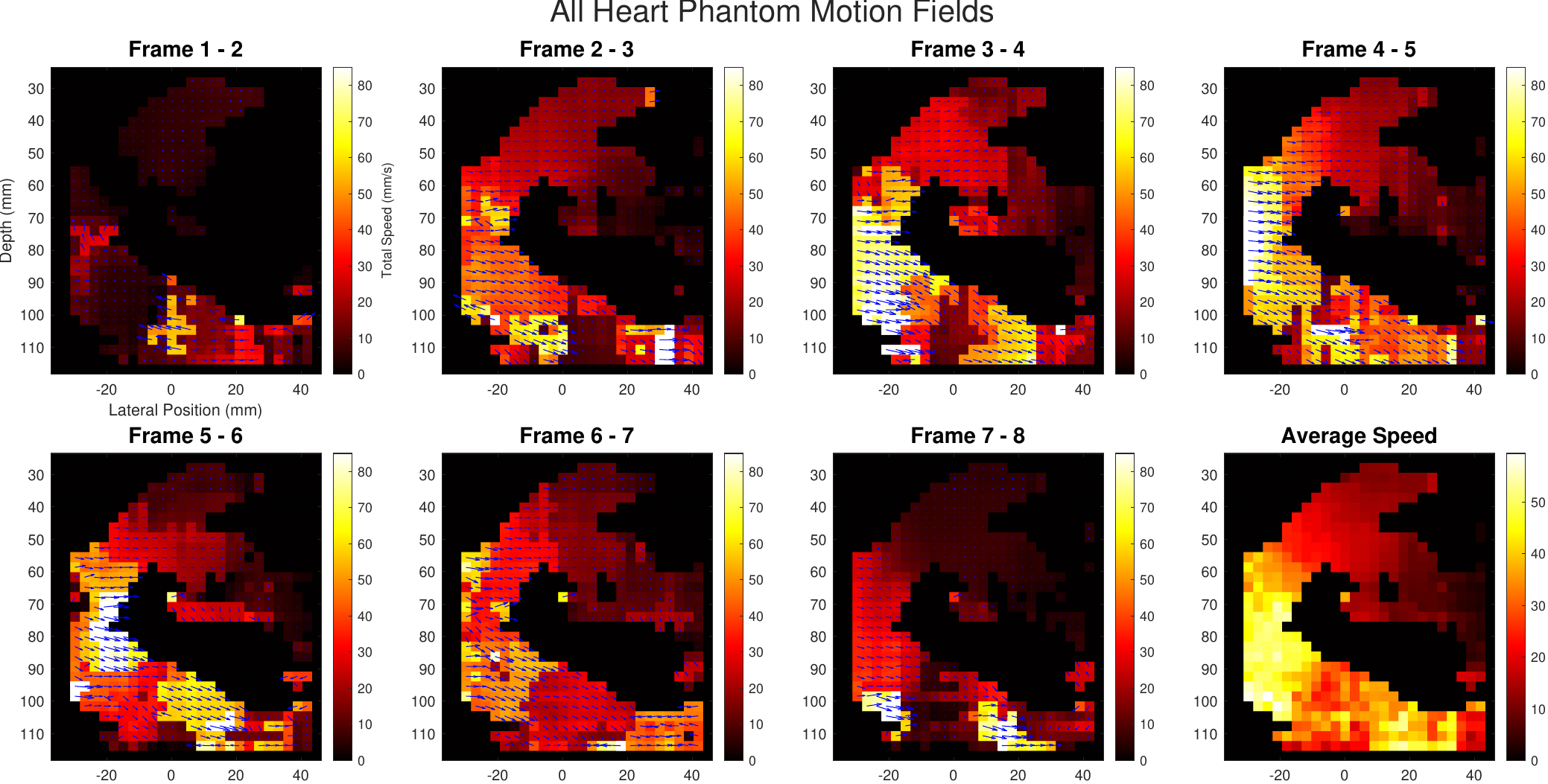}
    \caption{Collection of all 7 motion fields detected by \emcTwo\ in the heart phantom experiment. Each field has been downsampled by a factor of 2 for visualization purposes. Background colour indicates the absolute value of the field in mm/s. The final image is the average of the 7 fields' magnitudes and has a lower colour scale ceiling.}
    \label{fig:all_heart_fields}
\end{figure*}

This section contains additional results from the heart phantom experiment. Figure \ref{fig:all_heart_lris} shows all 8 \acs{readi} low-resolution images from the experiment, which are the inputs into \emcTwo\ and produce the results in Figures 8 and 9. Figure \ref{fig:all_heart_fields} shows all 7 motion fields detected by \emcTwo\ in this experiment, along with an average of their magnitudes.

\end{document}